\documentclass[reprint, superscriptaddress, aps, prx]{revtex4-2}

\usepackage{amsmath,amssymb}
\usepackage{amsthm}
\usepackage{physics}
\usepackage{graphicx}
\usepackage{hyperref}
\usepackage[dvipsnames]{xcolor}
\usepackage{tikz}
\usetikzlibrary{positioning,calc}
\usepackage{float}
\usepackage{booktabs}
\usepackage{algorithm}
\usepackage{algpseudocode} 
\usepackage{xparse}
\newcommand{\ZT}{\hat{\mathrm{zT}}} 
\newcommand{\DT}{\hat{\mathrm{DT}}} 
\newcommand{\QFT}{\hat{\mathrm{QFT}}} 
\newcommand{\SWAP}{\hat{\mathrm{SWAP}}} 

\newcommand{\im}{{\rm i}}

\definecolor{NJColor}{RGB}{0,120,0}
\definecolor{DPColor}{RGB}{0,70,180}
\definecolor{SRCColor}{RGB}{180,0,120}
\newif\ifshowchanges
\showchangestrue

\newsavebox{\deletedwordbox}
\newcommand{\deletedword}[1]{%
  \sbox{\deletedwordbox}{#1}%
  \rlap{\raisebox{0.45ex}[0pt][0pt]{\rule{\wd\deletedwordbox}{0.35pt}}}#1}
\ExplSyntaxOn
\NewDocumentCommand{\strikewords}{+m}{%
  \seq_set_split:Nnn \l_tmpa_seq { ~ } {#1}%
  \seq_map_inline:Nn \l_tmpa_seq { \deletedword{##1}\space }%
}
\ExplSyntaxOff

\newcommand{\smt}{Science, Mathematics and Technology Cluster, Singapore University of Technology and Design, 8 Somapah Road, 487372 Singapore}
\newcommand{\epd}{Engineering Product Development Pillar, Singapore University of Technology and Design, 8 Somapah Road, 487372 Singapore}
\newcommand{\cqt}{Centre for Quantum Technologies, National University of Singapore 117543, Singapore}

\begin{document}

\title{Quantum-Inspired Algorithms beyond Unitary Circuits: the Laplace Transform} 
\author{Noufal Jaseem}
\affiliation{\smt} 
\affiliation{\cqt}
\author{Sergi Ramos-Calderer} 
\affiliation{\cqt}
\author{Gauthameshwar S.}
\affiliation{\smt}
\affiliation{\cqt}
\author{Dingzu Wang} 
\affiliation{\smt} 
\affiliation{\cqt}
\author{Jos\'e Ignacio Latorre}
\affiliation{\cqt} 
\author{Dario Poletti}
\email{dario\_poletti@sutd.edu.sg}
\affiliation{\smt}
\affiliation{\epd}
\affiliation{\cqt} 
\date{\today}

\begin{abstract} 
Quantum-inspired algorithms can deliver substantial speedups over classical state-of-the-art methods by executing quantum algorithms with tensor networks on conventional hardware. Unlike circuit models restricted to unitary gates, tensor networks naturally accommodate non-unitary maps. 
This leads us to introduce a framework whereby one can design better performing algorithms with modular components inheriting a quantum algorithmic structure, yet going beyond unitarity thanks to the explicit use of non-unitary gates. 
As a concrete example, here we introduce a tensor-network approach to compute the discrete Laplace transform, a non-unitary, aperiodic transform (in contrast to the Fourier transform). We encode a length-$N$ signal on two paired $n$-qubit registers and decompose the overall map into a non-unitary exponential Damping Transform followed by a Quantum Fourier Transform. 
We prove that this decomposition admits strong MPO compression to low bond dimension resulting in significant acceleration, even for input data that cannot be compressed efficiently into matrix product states. 
We demonstrate simulations up to $N=2^{30}$ input data points, with up to $2^{60}$ output data points, and quantify how bond dimension controls runtime and accuracy, including precise and efficient inference of pole locations. 
\end{abstract}
\maketitle

{\it Introduction --- }Quantum algorithms have attracted significant attention in the past few decades as they can lead to exponential or polynomial speedup for certain computations \cite{grover1996fast,lloyd1996universal,shor1999polynomial,montanaro2016quantum}.     
However, their implementation requires fault tolerant quantum computation which is still out of reach of current technologies, and some applications lack a method of efficiently loading the data on the quantum processor  
\cite{preskill1998reliable,fowler2012surface,preskill2018quantum,aaronson2015read,biamonte2017quantum}. 
At the same time, up to certain sizes, it is possible to implement efficient simulations of some of these quantum algorithms, producing novel algorithms which run on classical machines~\cite{ma2022low,pan2022simulation,chen2023quantum,pan2024efficient,de2025dynamical,niedermeier2024simulating, GuoWu2019, zhuPan2021}.  

These are known as \emph{quantum-inspired} solutions, and while their realm of applicability is not as general as quantum algorithms, they provide faster algorithms in a relevant subset of cases. 
An important example of this strategy is the tensor-network implementation of the quantum Fourier transform \cite{chen2023quantum,niedermeier2024simulating, oseledets2010tt,dolgov2012qttfft}, that results in an acceleration over the fast Fourier transform on data efficiently compressible into tensor trains. 

The Fourier transform is a unitary transformation, ubiquitous in signal processing, that can be translated to an efficient quantum algorithm, as its periodic structure gives rise to an exponential speedup. 
However, mathematics has gifted us with other functional transforms that do not have a straightforward quantum analogue, including the Laplace transform. The Laplace transform is an extension of the Fourier transform to complex-valued frequencies, and is nowadays commonly used to solve linear differential equations~\cite{hayes1996statistical}.  
The discretized Laplace transform, also known as $z$-transform, is an important tool in signal processing, as it provides a complex-plane view of discrete signals, revealing stability, resonances, and decay properties. It has applications in digital filter design, systems' control, stability analysis etc. \cite{rabiner1969chirp,oppenheim1997signals,bozkurt2005zeros,tong1999rotation}.   
On the real positive axis, fast algorithms for the $z$-transform of a data-set of $N$ inputs evaluated at $M$ outputs achieve near-linear complexity $O(N+M)$~\cite{rokhlin1988fast,strain1992fast,loh2023fast}. More generally, when evaluating along a specific contour in the complex plane, the $z$-transform can be implemented via the chirp-$z$ approach with a computational cost of \(O\big((N+M)\log(N+M)\big)\) \cite{rabiner1969chirp,oppenheim1997signals}, 
while on a dense 2D concentric grid of possible outputs in the complex plane, with e.g. $M=N^2$, one would have a typical scaling of \(O\big(N^2\log(N)\big)\).

Crucially, the $z$-transform is not unitary, nor has the periodic properties of the Fourier transform. A common way to still retain a quantum-circuit representation, including in a classical simulation, is to embed the map in a larger unitary system and then project or trace out ancillas; recent quantum Laplace algorithms use such an encoding~\cite{zylberman2024fastlaplacetransformsquantum,singh2025}. 
Tensor networks, though, are not limited to unitary operations. 
One can thus consider a class of quantum-inspired algorithms which, while being structurally inspired by quantum algorithms in their modules, use operations beyond those available on a quantum processor.
As an example, our tensor-network calculation uses the QFT circuit layout as a guide but inserts the non-unitary damping maps directly as local tensors, without extra ancillas or a larger unitary embedding. 
This is our new conceptual step. Removing unitarity as a design restriction opens this circuit-guided MPO method to non-unitary classical transforms, and can be applied to multiple types on transform tapping on the modularity of the gates. 

\begin{figure*}[t]
    \includegraphics[width=0.98\linewidth]{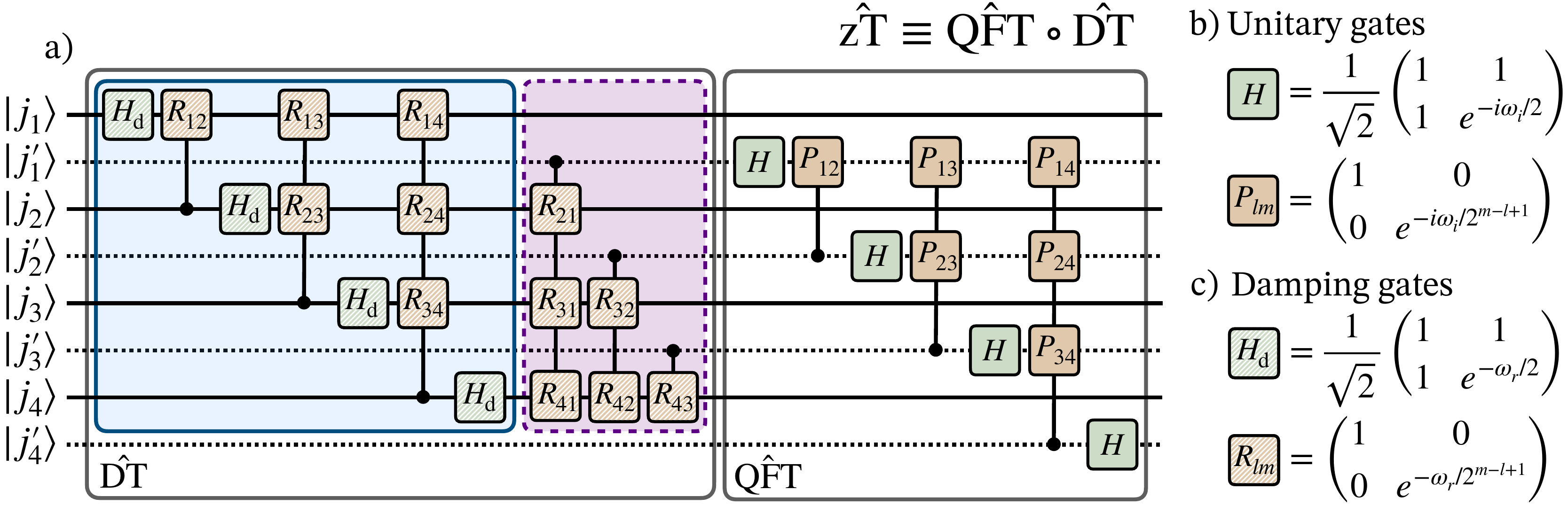}
  \caption{(a) Decomposition of the z-transform into the damping transform $\DT$, and the quantum Fourier transform $\QFT$, shown in the paired-register layout $\ket{j}\ket{j'}$. The $\DT$ consists of damping operations, defined in (c), separated into two steps. The solid blue box are damping operations acting on register $\ket{j}$, while the dashed purple box acts on both $\ket{j}\ket{j'}$, using register $\ket{j'}$ as controls. The $\QFT$ acts only on register $\ket{j'}$ and uses unitary gates shown in (b). The controlled operations can be understood as sequential applications of controlled quantum gates, or as a joint MPO where identity operations are applied on wire crossings.
  }
  \label{fig:zT_TN_diagram}
\end{figure*}

In this letter, we develop a new class of \emph{quantum-inspired} algorithms which, while building on quantum algorithms, they are assembled with non-unitary gates compressed into tensor networks. We thus formulate the \(z\)-transform ($\ZT$) 
as a low-bond-dimension tensor network acting on data encoded on a paired configuration of two registers $\ket{j}\ket{j}$. That is, only states $\ket{j}\ket{j'}$ with $j= j'$ are supported. 
The key observation is that on a polar grid $z_{k,\ell}=r_k e^{\im \theta_\ell}$, each sample weight in the $z$-transform factors as $z_{k,\ell}^{j}=r_k^{j} e^{\im j\theta_{\ell}}$, letting the radial and angular $j$-dependencies be realized on separate registers. This motivates a matrix-product-operator (MPO) realization composed of two main parts: a non-unitary \emph{Damping Transform} ($\DT$) implementing the radial factors $r_k^{j}$, and a \emph{Quantum Fourier Transform} ($\QFT$) implementing the phases $e^{\im j\theta_{\ell}}$. 
At fixed accuracy, we show that the maximum bond dimension of the tensor train singular value decomposition (TT-SVD) damping MPO is bounded by a polylogarithmic function of $N$~\cite{supp}. 
We also show that this leads to a computational complexity that scales, in the worst-case scenario of incompressible data, as $O\left( N^{3/2} \right)$ up to polylogarithmic corrections, which is favorable, on fine grids with $M=N^2$ output points in the complex plane, against current available methods. 
In scenarios in which the input can be compressed as an MPS, then the advantage is even more striking. 
We exemplify this scalable computation for data sizes up to $N=2^{30}$ inputs and $M=2^{60}$ outputs, showing that for input data that can be compressed as matrix product states (MPS), our method is even more advantageous. 
We also introduce a practical workflow for inferring pole locations from dense \(z\)-plane samples, with an adjustable resolution. 
All of the results shown here have been performed simply on a laptop.

{\it The $z$-transform --- }The $z$-transform of a finite sequence $\{x_j\}_{j=0}^{N-1}$ is
\begin{equation}
  \chi(z)=\sum_{j=0}^{N-1} x_j\,z^{j}, \qquad z=e^{-s},
\end{equation}
where $s=\sigma+\im\omega$ is the logarithmic coordinate on the $z$–plane ($\sigma=-\log|z|$, $\omega=-\arg z$), and $\im$ is the imaginary unit. We evaluate $\chi$ on the grid
 \( s_{k,\ell}={\left(\omega_r k+\im\,\omega_i \ell\right)}/{N},\; k,\ell=0,\dots,N-1\),
so that the real scales $\omega_r$ and $\omega_i$ set, respectively, the radial (damped) and angular (oscillatory) resolution and, simultaneously, the accessible region via $|z|=e^{-\omega_r k/N}$ and $\arg z=-\omega_i \ell/N$. 
The corresponding outputs are thus 
\begin{equation} \label{eq:chi_kl}
  \chi_{k,\ell}=\sum_{j=0}^{N-1} x_j\,e^{-(\omega_r k/N) j}\,e^{-\im(\omega_i \ell/N) j}.
\end{equation}
In what follows, we set $\omega_i=2\pi$, while $\omega_r$ remains tunable and controls the radial range and resolution.  Setting $\omega_r=0$ restricts to the unit circle (pure spectral content); $\omega_r>0$ shifts sampling inward ($|z|<1$), while $\omega_r<0$ shifts it outward ($|z|>1$). Throughout, we focus on $|z|\le 1$, where the weighted terms decay with $j$, yielding numerically well-conditioned evaluations.


\begin{figure*}[t]
    \includegraphics[width=0.98\linewidth]{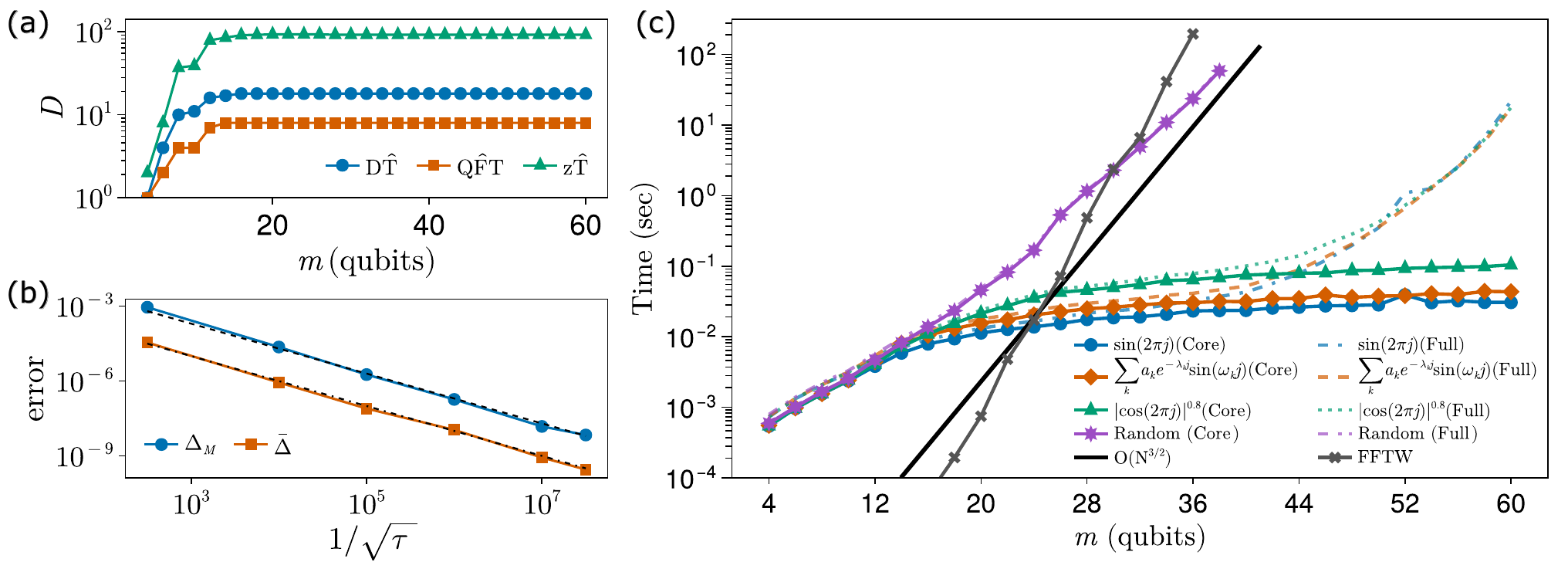} 
  \caption{ \textbf{Numerical results for the MPO representation of the $z$-Transform.} (a) Maximum MPO bond dimension $D$ for $\DT$, $\QFT$, and the composite $\ZT$ versus the number of output qubits $m = 2n=2\log_2(N)$ at cutoff $10^{-15}$ and $(\omega_r,\omega_i)=(2\pi,2\pi)$, indicating strong MPO compressibility. $N$ is the size of input data and $n$ is the number of input qubits. (b) Error versus the inverse square root of the SVD cutoff, $1/\sqrt{\tau}$, on log–log axes for $m=20$ with a Gaussian random input $x_j \sim \mathcal{N}(0,1)$. 
  Max and mean absolute errors, $\Delta_M$ and $\bar{\Delta}$ respectively, exhibit $\sqrt{\tau}$ scaling as confirmed by linear fits (dashed/dot–dashed). (c) 
  Transform-application runtime $t$ for the $z$-transform, implemented by sequential application of the $\DT$ MPO followed by the $\QFT$ MPO, with truncation cutoff $\tau=10^{-15}$, versus system size $m$. 
  “Full” includes the time required for encoding the signal into the initial MPS; “Core” does not include that cost. 
  Curves for several input types are shown (see legend and \cite{supp}). The solid black guide shows the expected worst-case scaling $t\propto N^{3/2}$. For the Gaussian random input, we also show the parallel radial-FFTW runtime on the same $N\times N$ polar grid. Over the displayed range, the FFTW runtime ($N^2\ln{N}$ scaling) exceeds the random-input MPO runtime. 
  }
  \label{fig:bond_error_runtime}
\end{figure*}


{\it Quantum-Inspired Representation --- }We encode a length-$N=2^n$ signal on two \emph{paired} $n$-qubit registers in a copy layout,
\begin{equation}\label{state_input}
  \ket{x}=\sum_{j=0}^{N-1} x_j\,\ket{j}\ket{j},\qquad
  \ket{j}=\ket{j_1 j_2\cdots j_n}.
\end{equation} 
Then, the (generally non-unitary) linear $z$–transform is the map $\ket{x}\mapsto \frac{1}{N} \sum_{k,\ell}\chi_{k,\ell}\ket{k}\ket{\ell}$. Equivalently, as an operator acting on the two registers,
\begin{equation}
  \ZT=\frac{1}{N} \sum_{j,k,\ell=0}^{N-1}
  e^{-(\omega_r k/N) j}\,e^{-\im(\omega_i \ell/N) j}\,
  \ket{k\ell}\!\bra{jj},
\end{equation}
where the $1/N$ in front of the sum is a rescaling factor. We implement $\ZT$ by a two-stage decomposition
\begin{equation}
  \ZT\equiv \QFT\circ \DT,
\end{equation}
where $\DT$ acts on register-1 with register-2 serving as a passive copy for control, and $\QFT$ acts on register-2. A schematic of this two-stage decomposition is depicted in Fig.~\ref{fig:zT_TN_diagram}(a). On computational basis states, it yields
\begin{align}
  \DT\,\ket{j}\ket{j'}
  &=\frac{1}{\sqrt{N}}\sum_{k=0}^{N-1} e^{-(\omega_r k/N) j}\,\ket{k}\ket{j'},\\[3pt]
  \QFT\,\ket{k}\ket{j'}
  &=\frac{1}{\sqrt{N}}\sum_{\ell=0}^{N-1} e^{-\im(\omega_i \ell/N) j'}\,\ket{k}\ket{\ell}. 
\end{align}
We write $\DT$ and $\QFT$ for the abstract maps above \footnote{We implicitly expect the data to be encoded into paired qubits register, this does not affect the scaling of the computational cost analysis.}; the circuit/MPO cores implement them up to bit-reversal of the output registers (final $\SWAP$), as in \cite{chen2023quantum}.

{\it Non-unitary gates and damping transform --- } 
The tensor network implementation of the quantum Fourier transform has been described in detail in \cite{chen2023quantum}.
Therefore, we focus here on the non-unitary gates used in the damping transform.       
We write $k=\sum_{\ell=1}^n k_\ell 2^{n-\ell}$ with $k_\ell\in\{0,1\}$, and thus, for a given $j$, we get 
 \( \sum_{k=0}^{N-1} e^{-\omega_r\,k\,j/N}\ket{k}
  = \bigotimes_{\ell=1}^{n} \left(\ket{0}+e^{-\omega_r\,j\,2^{-\ell}}\ket{1}\right)\),
which motivates defining a local ``damping-Hadamard'' gate  
\begin{eqnarray}
  \mathcal{H}_\mathrm{damp}
  &=& \frac{1}{\sqrt{2}}
  \begin{pmatrix}
    1 & 1 \\
    1 & e^{-\omega_r/2}
  \end{pmatrix}, 
\end{eqnarray}
where the prefactor $1/\sqrt{2}$ is only used as analogy with the Hadamard gate. 
Similarly, we also define ``controlled-damping'' gates which, for target qubit $\ell$ and control qubit $m$, act as 
\begin{equation}
  R_{l,m} =
  \exp\!\Big[-\theta_{l,m}\,\ketbra{1}{1}_l\otimes\ketbra{1}{1}_m\Big],
\end{equation} 
where $ \theta_{l,m} = \omega_r/2^{m-l+1} $. 
Importantly, unlike the 
controlled phases of the $\QFT$ which only need $m>l$ due to the $2\pi$ periodicity, the real exponent lacks periodicity, so the $\DT$ requires controls from \emph{all} bits $m$, both $m>l$ and $m<l$. 
We thus encode the data on two registers where the information appears only on matched pairs \(\ket{j}\ket{j}\), and the second register is used to implement the controls of the $\DT$ for $m<l$. 
The depiction of the $z$-transform in Fig.~\ref{fig:zT_TN_diagram} a) accommodates both a circuit and tensor network interpretation. In Fig.~\ref{fig:zT_TN_diagram} b) and c), the unitary and damping gates used in the graphical representation of the $\QFT$ and $\DT$ are defined. As a circuit, the controlled phase and damping gates are performed sequentially, while as a tensor network they can be merged together as a single unit. In this representation the horizontal lines correspond to physical indices, while the vertical ones are auxiliary legs whose size is given by the bond dimension. In this graphical representation, we understand a crossing between a horizontal and a vertical line as acting with an identity operation on that qubit. 
Importantly, to reduce the bond dimension required to represent the data as a matrix product state, we order the qubits by alternating those from each register. 
In summary, the algorithm is initialized with a matrix product state representation of the $N=2^n$ sized data encoded on matching pairs $\ket{j}\ket{j}$, then the matrix product operators corresponding to $\DT$ and $\QFT$  are applied sequentially. At last, a $\SWAP$ operation is performed, as done for the $\QFT$ algorithm. The first output register labels the radial index $k$, while the second labels the angular index $\ell$; their joint amplitudes encode
$\chi_{k,\ell}$. A detailed, step-by-step description of this algorithm can be found in \cite{supp}. %

{\it Results --- } Here we analyze the performance of this quantum-inspired strategy to perform the $z$-transform. 
Figure~\ref{fig:bond_error_runtime} a) shows the maximum MPO bond dimension reached for the $\DT$, the $\QFT$, and their composition $\ZT=\QFT\!\circ\!\DT$, evaluated at a fixed relative SVD cutoff $\tau=10^{-15}$. While it has been shown that the $\QFT$ can be represented efficiently (excluding the final $\SWAP$ operation) as an MPO in \cite{chen2023quantum}, we can show, 
in a fundamentally different way,
that the bond dimension for the $\DT$ is also under control. 

It is possible to prove, and we give the details in \cite{supp}, that the bond dimension of the TT-SVD MPO, for a given global accuracy and $\omega_r>0$, is upper bounded by 
\begin{equation}
 D_{\DT}^{(\epsilon)}
 \le
 C(\omega_r)\left[ 1+\log\frac{1}{\epsilon} +\frac{1}{2}\log\!\bigl(2\log_2 N-1\bigr) \right]^2,
 \label{eq:main_fixed_omega_bound}
\end{equation}
which implies $D_{\DT}^{(\epsilon)}
 = O_{\omega_r,\epsilon}\!\left[ \left( \log\log N \right)^2 \right]$ scaling. For the QFT core, the error bound of \cite{chen2023quantum} gives a conservative scaling $D_{\mathrm{QFT}}^{(\epsilon)} =O_{\epsilon}(\log\log N)$, which grows slowly with $N$ than $D_{\mathrm{DT}}^{(\epsilon)}$. Therefore, the $\DT$ contribution determines the leading polylogarithmic cost in a sequential application. 
Considering that the computational time complexity scales as $O[\chi^3 D^3 \log(N)]$, where $\chi$ is the bond dimension of the input MPS, $D$ the bond dimension of the MPO and $\log(N)$ is the size of the MPS, and considering that in the worst case scenario (e.g. random input data) $\chi\sim \sqrt{N}$, then the MPO-MPS application cost scales as $O[N^{3/2} \log(N) (\log\log N)^6]$, while the preparation of the initial MPS would require, at most, $O[N^{3/2} (\log(N))]$. 
It is important to note that the prefactor that depends on $\omega_r$ can grow significantly as $\omega_r\rightarrow 0$ or it is large. We can, however, derive an $\omega_r$ independent bound which goes as 
\begin{equation}
 D_{\DT}^{(\epsilon)}
\le C_* \left\{ (1+\log N)
\left[
1+\log N+\log\!\left(\frac{1}{\epsilon}\right)
\right] \right\}^{2}, 
\end{equation} 
which implies that for a given $\epsilon$ it goes as $[\log(N)]^4$. Overall, apart from different polylogarithmic terms, we thus get the same scalings as for the $\omega_r$ dependent bound. More details can be found in~\cite{supp}. 

We now consider some numerical examples. 
With an MPO representation of the $z$-transform, we can choose a different relative cut-off point $\tau$ for the singular values which affect the accuracy of the $z-$transform. To do so, we consider a signal of length $N=2^{10}$ with $x_j$ drawn independently from a zero-mean, unit-variance Gaussian ($x_j \sim \mathcal{N}(0,1)$), so that the transform can be readily computed exactly. 
We calculate an input–$\ell_1$–normalized per-sample absolute error
\(\Delta_{k,\ell}
=|\left(\tilde{\chi}_{k,\ell}-\chi_{k,\ell}\right)|/{\sum_{j=0}^{N-1}|x_j|},
\,{\rm with \;}
\tilde{\chi}_{k,\ell}:=N\,\langle k,\ell \,|\, \ZT \,|\, x\rangle,\)
where $\chi_{k,\ell}$ denotes the analytical value (Eq.~\ref{eq:chi_kl}), and $\tilde{\chi}_{k,\ell}$ is obtained from the MPO $\ZT$ compressed with relative cut-off $\tau$. We report the mean, $\bar{\Delta}$ and maximum, $\Delta_M$ of $\Delta_{k,\ell}$ over the sampled $(k,\ell)$ grid. Figure~\ref{fig:bond_error_runtime}b) shows both metrics versus $\tau$.
From numerical fits, both errors scale approximately as $\sqrt{\tau}$, with $\Delta_M\approx 0.2 \sqrt{\tau}$ and $\bar{\Delta}\approx 10^{-2}\sqrt{\tau}$.   

\begin{figure}[t!]
\includegraphics[width=0.99\columnwidth]{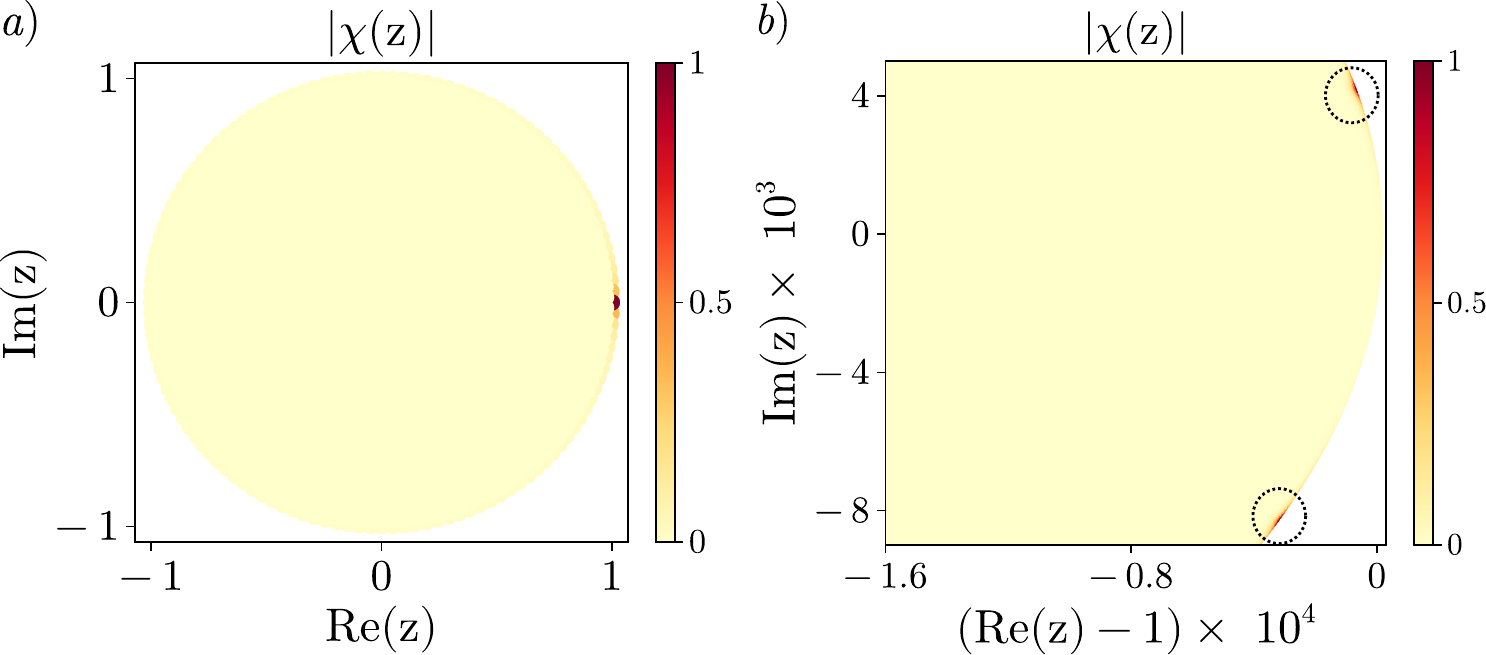}
\caption{Pole-location inference from the compressed MPS at $n=20$, accessing $2^{40}$ candidate samples without materializing the full grid. Color encodes normalized $|\chi|$ ($|\chi|/\max|\chi|$).  Each sample is computed on a $(k,\ell)$ grid point and plotted at the corresponding $z$-plane location $z_{k,\ell}=e^{-s_{k,\ell}}$. 
(a) $|\chi(z)|$ on a coarse $(k\ell)$ grid (step size $2^{12}$; $256\times256$ samples), with the MPO defined by $\omega_r= \omega_i = 2\pi$, reveals a peak near $z\approx 1$. 
(b) A fine scan, carried out with a different MPO using $\omega_r=0.5$, $\omega_i = 2\pi$ for improved radial resolution, centered at $z=1$ resolves two peaks at $z \approx 0.99997+0.00409\,\im$ and $z \approx 0.99994-0.00817\,\im$. Across the plotted samples, the max absolute error, $ \Delta_M\lesssim 10^{-7}$. The extraction of the corresponding pole locations from the finite-\(N\) data is described in \cite{supp}. 
} 
  \label{fig:poles}
\end{figure}

We have established that the MPO representation of the $z$-transform is accurate. We now study its efficiency.
Figure~\ref{fig:bond_error_runtime}(c) shows end-to-end runtimes for different types of input data, depicted by different symbols.
For each case, we show both the time needed to execute only the $\ZT$, referred to as \emph{Core} and depicted by continuous lines, and the total runtime, including the additional time required to convert the data into a matrix product state, referred to as \emph{Full} and depicted by dashed lines. The data MPS is generated using randomized SVD with a truncation error threshold of $10^{-15}$. For input data, we use a single sinusoidal function (blue circles), a sum of 10 sinusoidal functions each with a different exponential decay (orange diamonds), the absolute value of a sinusoidal function (green triangles), and random numbers from a normal distribution (purple stars).
More details are given in \cite{supp}. 
The runtime for non-random data, benchmarked here up to $N=2^{30}$, 
is lower than the random-input runtime over the tested sizes. 
For random input, tested up to $N=2^{19}$, and we compare this to the $N^{3/2}$ line in Fig.~\ref{fig:bond_error_runtime}(c) which is a guide of the asymptotic scaling for this case. 
For comparison with a conventional FFT-based implementations, we evaluate the $z$-transform on the same $N\times N$ polar grid used for the MPO calculation. For each of the $N$ radii, we form the damped signal $x_j\exp(-\omega_r kj/N)$ and use a length-$N$ FFT to obtain all $N$ angular samples, giving an overall computational cost of $O(N^2\log N)$. Similar scaling is obtained with the chirp-$z$ transform. 
Instead, whether we consider a finite $\omega_r$, or an $\omega_r$ independent analysis, our methods scales, at fixed $\epsilon$, as $O(N^{3/2})$ up to logarithmic corrections, and thus better than $O(N^2)$ on the full grid. 

At this point, considering the large size of the output data, an important matter to tackle is the accessibility of the data. 
While all the information is efficiently stored in the output MPS, our approach is not advantageous if one aims to extract each single output value. Instead, it is very advantageous in two scenarios: first, when the output MPS can also be passed directly to subsequent tensor-network operations without materializing all $M=N^2$ values. This is useful when the transform is an intermediate step in a larger calculation, as is often the case for Fourier transforms; second, when one analyzes the output in coarse grained manner before zooming in the most relevant part.          
We give here an example. 
We consider as input data the sequence $x_j \propto a^{j}\cos(\omega_{0} j)$, whose $z$-transform for an infinite series has poles at $z=(1/a)\,e^{\pm \im\omega_{0}} \approx 0.99984 + 0.00408\,\im,\ 0.99981 - 0.00816\,\im$, for $a=1.00015 e^{\im 0.002}$ and $\omega_{0} = 0.0061$. These values are rounded for readability; see \cite{supp} for details. 
In Fig.~\ref{fig:poles}(a) we show the coarse-grained form of the $z-$transform, whereby we only consider the first $8$ most significant bits of each output index $k$ and $\ell$.    
%
This shows that the $z$-transform has, at least, a pole close to the $z=1$ point in the complex plane. The MPS representation allows us to zoom in to that portion of the complex plane and clearly single out the two inferred poles locations, as shown in Fig.~\ref{fig:poles}(b), see \cite{supp}. 
Alternatively, we can also use the autoregressive properties of the solution matrix product state to sample the node configurations. 
The same scanning strategy can also be applied to find zeros. An example is presented in \cite{supp}, 
where three closely spaced zeros are clearly resolved in the finite-\(N\) truncated transform as localized minima of \(|X(z)|\).

{\it Conclusions --- } 
We introduced a new paradigm of quantum-inspired computing that also uses non-unitary gates as building blocks. 
We have exemplified this with a realization of the 
quantum-inspired discrete Laplace transform, or $z$-transform. 
We have proven that our quantum-inspired tensor network implementation of the $z$-transform offers efficient computation for signals of large input size, and dense sampling in the z-plane, even for random input data which cannot be efficiently compressed. 
We have represented the map $x_j \mapsto \chi_{k,\ell}$ as a compressed MPO acting on a signal encoded into a matrix-product state which returns a matrix product state over the output indices $(k,\ell)$. This enables on-demand evaluation of dense polar grids and multiple contours without materializing the full array with $M$ elements. 
We have shown that for MPS-compressible input data there can be a speed up of various orders of magnitude compared to standard approaches, and we have demonstrated reliable poles (and zeros) location inference on large grids.  
This can lead to applications, for instance in the design of transmission functions, whose properties depend significantly on the location of their zeros and poles. 
More importantly, the modular structure of our algorithm, and the ability to use both unitary and non-unitary gates, can lead to realization of other non-unitary operations such as the Hankel and Mellin transforms, Gaussian kernels and more. Whether these cases admit efficient constructions and compression bounds remains an open question.   
Our implementation is available as open-source code on GitHub~\cite{QILaplacejl}. 


\vspace{4pt}
\noindent\textbf{Acknowledgements.} We acknowledge support from the Singapore Ministry of Education from grant MOE-T2EP50123-0017, and from the Centre for Quantum Technologies grant CQT$\_$SUTD$\_$2025$\_$01. 
Constructing the $z$-transform MPO and related circuit tensors was performed using the software package ITensor \cite{fishman2022itensor} in Julia.
\bibliographystyle{apsrev4-2}
\bibliography{bib_file}

@article{chen2023quantum,
  title   = {Quantum Fourier Transform Has Small Entanglement},
  author  = {Chen, Jielun and Stoudenmire, E.M. and White, Steven R.},
  journal = {PRX Quantum},
  volume  = {4},
  pages   = {040318},
  year    = {2023},
  doi     = {10.1103/PRXQuantum.4.040318},
  url     = {https://link.aps.org/doi/10.1103/PRXQuantum.4.040318}
}

@book{oppenheim1997signals,
  title     = {Signals and Systems},
  author    = {Oppenheim, Alan V. and Willsky, Alan S. and Nawab, S. Hamid},
  edition   = {2nd},
  publisher = {Prentice Hall},
  url={https://www.pearson.com/en-us/subject-catalog/p/signals-and-systems/P200000003155/9780138147570},
  year      = {1997}
}

@article{tong1999rotation,
  title={Rotation of NMR images using the 2D chirp-z transform},
  author={Tong, Raoqiong and Cox, Robert W},
  journal={Magnetic Resonance in Medicine: An Official Journal of the International Society for Magnetic Resonance in Medicine},
  volume={41},
  number={2},
  pages={253--256},
  year={1999},
  url={https://sscc.nimh.nih.gov/sscc/rwcox/papers/Chirpz1999.pdf},
  publisher={Wiley Online Library}
}

@article{bozkurt2005zeros,
  title={Zeros of z-transform representation with application to source-filter separation in speech},
  author={Bozkurt, Baris and Doval, Boris and d'Alessandro, Christophe and Dutoit, Thierry},
  journal={IEEE signal processing letters},
  volume={12},
  number={4},
  pages={344--347},
  year={2005},
  doi={10.1109/lsp.2005.843770},
  publisher={IEEE}
}

@misc{zylberman2024fastlaplacetransformsquantum,
      title={Fast Laplace transforms on quantum computers}, 
      author={Julien Zylberman},
      year={2024},
      eprint={2412.05173},
      archivePrefix={arXiv},
      primaryClass={quant-ph},
      url={https://arxiv.org/abs/2412.05173}, 
}

@article{fishman2022itensor,
  title={The ITensor software library for tensor network calculations},
  author={Fishman, Matthew and White, Steven and Stoudenmire, Edwin Miles},
  journal={SciPost Physics Codebases},
  pages={004},
  doi={10.21468/SciPostPhysCodeb.4},
  year={2022}
}

@article{rabiner1969chirp,
  title={The chirp z-transform algorithm and its application},
  author={Rabiner, Lawrence R and Schafer, Ronald W and Rader, Charles M},
  journal={Bell System Technical Journal},
  volume={48},
  number={5},
  pages={1249--1292},
  year={1969},
  doi={10.1002/j.1538-7305.1969.tb04268.x},
  publisher={Wiley Online Library}
}

@article{shor1999polynomial,
  title={Polynomial-time algorithms for prime factorization and discrete logarithms on a quantum computer},
  author={Shor, Peter W},
  journal={SIAM review},
  volume={41},
  number={2},
  pages={303--332},
  year={1999},
  doi= {10.1137/S0036144598347011},
  publisher={SIAM}
}

@inproceedings{grover1996fast,
  title={A fast quantum mechanical algorithm for database search},
  author={Grover, Lov K},
  booktitle={Proceedings of the twenty-eighth annual ACM symposium on Theory of computing},
  pages={212--219},
  year={1996},
  doi = {10.1145/237814.237866}
}

@article{lloyd1996universal,
  title={Universal quantum simulators},
  author={Lloyd, Seth},
  journal={Science},
  volume={273},
  number={5278},
  pages={1073--1078},
  year={1996},
  doi = {10.1126/science.273.5278.1073},
  publisher={American Association for the Advancement of Science}
}

@article{montanaro2016quantum,
  title={Quantum algorithms: an overview},
  author={Montanaro, Ashley},
  journal={npj Quantum Information},
  volume={2},
  number={1},
  pages={1--8},
  year={2016},
  doi={10.1038/npjqi.2015.23},
  publisher={Nature Publishing Group}
}

@article{preskill1998reliable,
  title={Reliable quantum computers},
  author={Preskill, John},
  journal={Proceedings of the Royal Society of London. Series A: Mathematical, Physical and Engineering Sciences},
  volume={454},
  number={1969},
  pages={385--410},
  year={1998},
  doi={10.1098/rspa.1998.0167},
  publisher={The Royal Society}
}

@article{fowler2012surface,
  title={Surface codes: Towards practical large-scale quantum computation},
  author={Fowler, Austin G and Mariantoni, Matteo and Martinis, John M and Cleland, Andrew N},
  journal={Physical Review A-Atomic, Molecular, and Optical Physics},
  volume={86},
  number={3},
  pages={032324},
  year={2012},
  doi= {10.1103/PhysRevA.86.032324},
  publisher={APS}
}

@article{preskill2018quantum,
  title={Quantum computing in the NISQ era and beyond},
  author={Preskill, John},
  journal={Quantum},
  volume={2},
  pages={79},
  year={2018},
  doi= {10.22331/q-2018-08-06-79},
  publisher={Verein zur F{\"o}rderung des Open Access Publizierens in den Quantenwissenschaften}
}

@article{aaronson2015read,
  title={Read the fine print},
  author={Aaronson, Scott},
  journal={Nature Physics},
  volume={11},
  number={4},
  pages={291--293},
  year={2015},
  doi={10.1038/nphys3272},
  publisher={Nature Publishing Group UK London}
}

@article{biamonte2017quantum,
  title={Quantum machine learning},
  author={Biamonte, Jacob and Wittek, Peter and Pancotti, Nicola and Rebentrost, Patrick and Wiebe, Nathan and Lloyd, Seth},
  journal={Nature},
  volume={549},
  number={7671},
  pages={195--202},
  year={2017},
  doi={10.1038/nature23474},
  publisher={Nature Publishing Group UK London}
}

@misc{supp,
  howpublished = {See Supplemental Material}
}

@article{rokhlin1988fast,
  title={A fast algorithm for the discrete Laplace transformation},
  author={Rokhlin, Vladimir},
  journal={Journal of Complexity},
  volume={4},
  number={1},
  pages={12--32},
  year={1988},
  url={https://doi.org/10.1016/0885-064X(88)90007-6},
  publisher={Elsevier}
}

@article{strain1992fast,
  title={A fast Laplace transform based on Laguerre functions},
  author={Strain, John},
  journal={mathematics of computation},
  volume={58},
  number={197},
  pages={275--283},
  doi={10.1090/s0025-5718-1992-1106983-2},
  year={1992}
}

@article{loh2023fast,
  title={Fast discrete Laplace transforms},
  author={Loh, Yen Lee},
  journal={Journal of Computational Mathematics and Data Science},
  volume={8},
  pages={100082},
  year={2023},
  doi = {https://doi.org/10.1016/j.jcmds.2023.100082},
  publisher={Elsevier}
}

@book{hayes1996statistical,
  title={Statistical digital signal processing and modeling},
  author={Hayes, Monson H},
  year={1996},
  url={https://www.wiley.com/en-us/Statistical+Digital+Signal+Processing+and+Modeling-p-9780471594314},
  publisher={John Wiley \& Sons}
}

@article{niedermeier2024simulating,
  title={Simulating the quantum Fourier transform, Grover's algorithm, and the quantum counting algorithm with limited entanglement using tensor networks},
  author={Niedermeier, Marcel and Lado, Jose L and Flindt, Christian},
  journal={Physical Review Research},
  volume={6},
  number={3},
  pages={033325},
  year={2024},
  doi = {10.1103/PhysRevResearch.6.033325},
  publisher={APS}
}

@article{de2025dynamical,
  title={Dynamical cluster-based strategy for improving tensor network algorithms in quantum circuit simulations},
  author={De Girolamo, Andrea and Facchi, Paolo and Rabl, Peter and Pascazio, Saverio and Lupo, Cosmo and Magnifico, Giuseppe},
  journal={Physical Review Research},
  volume={7},
  number={4},
  pages={043170},
  year={2025},
  doi = {10.1103/x7x4-kn4n},
  publisher={APS}
}

@article{ma2022low,
  title={Low rank approximation in simulations of quantum algorithms},
  author={Ma, Linjian and Yang, Chao},
  journal={Journal of Computational Science},
  volume={59},
  pages={101561},
  year={2022}, 
  url={https://doi.org/10.1016/j.jocs.2022.101561}, 
  publisher={Elsevier}
}

@article{pan2022simulation,
  title={Simulation of quantum circuits using the big-batch tensor network method},
  author={Pan, Feng and Zhang, Pan},
  journal={Physical Review Letters},
  volume={128},
  number={3},
  pages={030501},
  year={2022},
  doi={10.1103/PhysRevLett.128.030501},
  publisher={APS}
}

@article{pan2024efficient,
  title={Efficient quantum circuit simulation by tensor network methods on modern gpus},
  author={Pan, Feng and Gu, Hanfeng and Kuang, Lvlin and Liu, Bing and Zhang, Pan},
  journal={ACM Transactions on Quantum Computing},
  volume={5},
  number={4},
  pages={1--26},
  year={2024},
  doi={10.1145/3696465},
  publisher={ACM New York, NY}
}

@misc{QILaplacejl,
  title        = {{QILaplace.jl}: Quantum-Inspired Laplace Transform},
  howpublished = {\url{https://github.com/SUTD-MDQS/QILaplace.jl}},
  note         = {GitHub repository, accessed 2026-01-07},
  year         = {2026}
}

@misc{singh2025,
      title={A Polylogarithmic-Time Quantum Algorithm for the Laplace Transform}, 
      author={Akash Kumar Singh and Ashish Kumar Patra and Anurag K. S. V. and Sai Shankar P. and Ruchika Bhat and Jaiganesh G},
      year={2025},
      eprint={2512.17980},
      archivePrefix={arXiv},
      primaryClass={quant-ph},
      url={https://arxiv.org/abs/2512.17980}, 
}

@article{GuoWu2019,
  title = {General-Purpose Quantum Circuit Simulator with Projected Entangled-Pair States and the Quantum Supremacy Frontier},
  author = {Guo, Chu and Liu, Yong and Xiong, Min and Xue, Shichuan and Fu, Xiang and Huang, Anqi and Qiang, Xiaogang and Xu, Ping and Liu, Junhua and Zheng, Shenggen and Huang, He-Liang and Deng, Mingtang and Poletti, Dario and Bao, Wan-Su and Wu, Junjie},
  journal = {Phys. Rev. Lett.},
  volume = {123},
  issue = {19},
  pages = {190501},
  numpages = {6},
  year = {2019},
  month = {Nov},
  publisher = {American Physical Society},
  doi = {10.1103/PhysRevLett.123.190501},
  url = {https://link.aps.org/doi/10.1103/PhysRevLett.123.190501}
}

@misc{zhuPan2021,
      title={Quantum Computational Advantage via 60-Qubit 24-Cycle Random Circuit Sampling}, 
      author={Chu Guo, et al.},
      year={2021},
      eprint={2109.03494},
      archivePrefix={arXiv},
      primaryClass={quant-ph},
      url={https://doi.org/10.48550/arXiv.2109.03494}, 
}

@article{oseledets2010tt,
  title   = {Approximation of {$2^d \times 2^d$} Matrices Using Tensor Decomposition},
  author  = {Oseledets, Ivan V.},
  journal = {SIAM Journal on Matrix Analysis and Applications},
  volume  = {31},
  number  = {4},
  pages   = {2130--2145},
  year    = {2010},
  doi     = {10.1137/090757861}
}

@article{dolgov2012qttfft,
  title   = {Superfast {F}ourier Transform Using {QTT} Approximation},
  author  = {Dolgov, Sergey and Khoromskij, Boris N. and Savostyanov, Dmitry V.},
  journal = {Journal of Fourier Analysis and Applications},
  volume  = {18},
  number  = {5},
  pages   = {915--953},
  year    = {2012},
  doi     = {10.1007/s00041-012-9227-4}
}

@misc{Koyama2023,
  title         = {Exponential Sum Approximations of Finite Completely Monotonic Functions},
  author        = {Koyama, Yohei M.},
  year          = {2023},
  eprint        = {2301.08931},
  archivePrefix = {arXiv},
  primaryClass  = {math.NA},
  doi           = {10.48550/arXiv.2301.08931},
  url           = {https://arxiv.org/abs/2301.08931}
}

@article{EckartYoung1936,
  title   = {The Approximation of One Matrix by Another of Lower Rank},
  author  = {Eckart, Carl and Young, Gale},
  journal = {Psychometrika},
  volume  = {1},
  number  = {3},
  pages   = {211--218},
  year    = {1936},
  doi     = {10.1007/BF02288367}
}

\end{document}


\title{Supplemental Material for\\
\textit{Quantum-Inspired Algorithms beyond Unitary Circuits: the Laplace Transform}}
\author{Noufal Jaseem}
\affiliation{\smt}
\affiliation{\cqt}
\author{Sergi Ramos-Calderer}
\affiliation{\cqt}
\author{Gauthameshwar S.}
\affiliation{\smt}
\affiliation{\cqt}
\author{Dingzu Wang}
\affiliation{\smt}
\affiliation{\cqt}
\author{Jos\'e Ignacio Latorre}
\affiliation{\cqt}
\author{Dario Poletti}
\email{dario\_poletti@sutd.edu.sg}
\affiliation{\smt}
\affiliation{\epd}
\affiliation{\cqt}
\date{\today}
\maketitle

\section{Damping transform}\label{app:DT}
Given that the application of tensor networks to implement the Quantum Fourier Transform (QFT) has been previously investigated by Chen, Stoudenmire, and White in their 2023 study~\cite{chen2023quantum}, this work narrows its focus to the implementation of the real-exponential component. For each $j$, the transformation can be expressed as follows.
\begin{equation}
    \ket{j}\ket{j'}
    \;\overset{\text{DT}}{\longmapsto}\;
    \frac{1}{\sqrt{2^n}}
    \sum_{k=0}^{2^n-1}
    \exp\Bigl(\frac{-\omega_r\,k\,j}{2^n}\Bigr)\,
    \ket{k} \;\ket{j'}.
    \label{eq_dt}
\end{equation}
Here, $\ket{j'}$ denotes the second register, an identical copy of $\ket{j}$, which is left unchanged by this step. Since the DT is nonunitary, its output vector need not be normalized. Write $k$ in binary as $k = k_1\,2^{n-1} + k_2\,2^{n-2} + \dots + k_n\,2^0$ (where each $k_\ell \in \{0,1\}$). Then
\begin{align}
   \sum_{k=0}^{2^n-1}& 
   e^{\frac{-\omega_r\,k\,j}{2^n}}\,
   \ket{k} =
   \sum_{k_1=0}^1 \sum_{k_2=0}^1 \cdots \sum_{k_n=0}^1
   \exp\Bigl[-\omega_r\,j
         \bigl(
           k_1\,2^{-1} \nonumber\\& \qquad\qquad+ k_2\,2^{-2} + \dots + k_n\,2^{-n}
         \bigr)\Bigr]\,
   \ket{k_1\,k_2\,\cdots\,k_n}\nonumber
   \\
   &=
   \sum_{k_1=0}^1 \cdots \sum_{k_n=0}^1 \prod_{\ell=1}^n
   \exp\Bigl(-\omega_r\,j\,k_\ell\,2^{-\ell}\Bigr)\,
   \ket{k_1\,k_2\,\cdots\,k_n}.
\end{align}
We can factorize the sum over each qubit index:
\begin{align}
   \Rightarrow & \frac{1}{\sqrt{2^n}}\sum_{k_1=0}^1 \cdots \sum_{k_n=0}^1
   \exp\Bigl(-\omega_r\,j\,(k_1 2^{-1} + \dots + k_n 2^{-n})\Bigr)\,\nonumber\\
   & \qquad\qquad\qquad\qquad\qquad\qquad\qquad\quad\ket{k_1\,\dots\,k_n} \ket{j'}\nonumber\\
   &=
   \frac{1}{\sqrt{2^n}}\bigl(\ket{0} + e^{-\omega_r\,j\,2^{-1}}\ket{1}\bigr)
   \;
   \bigl(\ket{0} + e^{-\omega_r\,j\,2^{-2}}\ket{1}\bigr)\nonumber\\
   & \qquad\qquad\qquad\qquad\quad
   \dots
   \bigl(\ket{0} + e^{-\omega_r\,j\,2^{-n}}\ket{1}\bigr) \ket{j'}.
   \label{eq_z_final}
\end{align}
Equation~\eqref{eq_z_final} gives the required DT output for one basis input $j$. By linearity, this defines the transform for any intended input $\sum_j x_j\ket{j}\ket{j}$. Later, we show that the gate circuit gives the same output, up to the final bit reversal. This verifies that the circuit implements the DT. Before describing the circuit, we show that the corresponding operator admits a low-bond-dimension MPO approximation.

\section{Bond-dimension bounds for the damping transform}
\label{sec:analytical_bound}

We first bound the approximate Schmidt rank of the scalar damping kernel in
Sec.~\ref{sec:dampingkernel}. At fixed $\omega_r>0$ and fixed accuracy at one cut, this
bound is independent of $N$; its accuracy dependence is derived explicitly
in Sec.~\ref{sec:o1_fixed_omega_accuracy}. We then extend the result to the
two-register damping operator and its MPO representation in
Sec.~\ref{sec:two_register_bound}. However, the prefactors in these
$N$-independent bounds become large as $\omega_r\to0^+$ or for very large
$\omega_r$. We therefore derive a second bound in Sec.~\ref{sec:koyama} that holds uniformly for all $\omega_r\geq 0$ and gives a bond dimension that is polylogarithmic in $N$.


\subsection{Preamble to the proofs}\label{sec:preamble}
We work with the no-SWAP circuit core, before the final bit reversal. Along the MPO chain, the input bits run from most to least significant. The first-register output bits run in reverse order, while the second register keeps the input ordering. For the scalar damping kernel, $\mathcal{D}_N(k,j)$, after expanding $k$ and $j$ in bits, we fuse the input-output bit pair $(j_i,k_i)$ into one local MPO site. For the two-register operator, $(h_i,k_i)$ and $(h'_i,k'_i)$ form two consecutive MPO sites. We refer to these two sites as a \emph{register pair}. With this convention, the cuts used below coincide with the bond cuts of the circuit MPO. Figure~\ref{fig:o1_proof_roadmap} summarizes the proof.

\begin{figure}[!t]
\centering
\resizebox{0.98\columnwidth}{!}{%
\begin{tikzpicture}[
  node distance=0.47cm,
  box/.style={rounded corners, draw=dtblue, fill=dtlight, thick, align=center,
              minimum width=3.25cm, minimum height=0.66cm, font=\scriptsize},
  wide/.style={box, minimum width=3.75cm},
  arr/.style={-{Latex[length=1.8mm]}, thick, draw=dtblue}
]
\node[wide] (cut) {Cut after $r$ kernel bit pairs\\
                    $k=k_L+Xk_R$, $j=Yj_L+j_R$};
\node[wide, below=of cut] (factor) {Separate the factors across the bond\\
                    left and right local factors $\times$ two left-right coupling kernels};
\node[wide, below=of factor] (cross) {Cross kernels\\
                    $P_{X,Y}=\e^{-\omega_r uv/N}$, $Q_{X,Y}=\e^{-\omega_r uv}$};

\node[box, below=0.62cm of cross, xshift=-2.12cm] (P) {$P$: bounded exponent\\
                    $0\leq \omega_r uv/N<\omega_r$};
\node[box, below=0.62cm of cross, xshift=2.12cm] (Q) {$Q$: fixed exponential decay\\
                    independent of $N$};
\node[box, below=of P] (cheb) {Chebyshev polynomial\\
                    degree $d-1$};
\node[box, below=of Q] (rows) {Keep the first $U+1$ rows\\
                    and one residual column};
\node[box, below=of cheb] (Prank) {$\operatorname{rank}(\widetilde P)\leq d$\\
                    entrywise relative error $\delta_d$};
\node[box, below=of rows] (Qrank) {$\operatorname{rank}(\widetilde Q)\leq U+2$\\
                    Frobenius error at most $\eta_U$};

\node[wide, below=0.66cm of $(Prank)!0.5!(Qrank)$] (combine) {Expand into product terms across the cut\\
                    $\operatorname{rank}\leq d(U+2)$};
\node[wide, below=of combine] (error) {Relative error\\
                    $\delta_d+(1+\delta_d)\eta_U$};
\node[wide, below=of error] (circuit) {Extend  the bound to the two-register damping MPO\\
                    same $P,Q$ at pair cuts; factor $\leq2$ inside a pair};
\node[wide, below=of circuit] (cutresult){For any fixed bond $r$, the bond dimension required\\ at tolerance $\eps_{\rm bond}$ satisfies the scaling\\ $D_r^{(\eps_{\rm bond})}=O_{\omega_r}(1)$};

\node[wide, below=of cutresult] (globalresult)
  {For a TT-SVD MPO with overall error tolerance $\eps$,\\ the maximum bond dimension satisfies the scaling\\ $D_{\mathrm{TT\text{-}SVD}}^{(\eps)} =O_{\omega_r,\eps}\!\left((\log\log N)^2\right)$};

\draw[arr] (cut) -- (factor);
\draw[arr] (factor) -- (cross);
\draw[arr] (cross) -- (P);
\draw[arr] (cross) -- (Q);
\draw[arr] (P) -- (cheb);
\draw[arr] (Q) -- (rows);
\draw[arr] (cheb) -- (Prank);
\draw[arr] (rows) -- (Qrank);
\draw[arr] (Prank) -- (combine);
\draw[arr] (Qrank) -- (combine);
\draw[arr] (combine) -- (error);
\draw[arr] (error) -- (circuit);
\draw[arr] (circuit) -- (cutresult);
\draw[arr] (cutresult) -- (globalresult);
\end{tikzpicture}%
}
\caption{Roadmap of the approximate-rank proof at fixed $\omega_r$. The bit split separates the damping kernel into one-sided factors and two crossing kernels. The crossing kernels are approximated separately and combined to bound the rank at each cut. The last two steps distinguish the single-bond bound from the TT-SVD bond-dimension bound at fixed overall error tolerance.}
\label{fig:o1_proof_roadmap}
\end{figure}

For a coefficient tensor $F$, let $F^{[r]}$ denote the matrix obtained by grouping all indices on either side of cut $r$ in the no-SWAP site ordering, and write
\begin{equation}
 D_r^{(\eps)}(F):=
 \min\left\{\operatorname{rank}\!\left(\widetilde F^{[r]}\right):
 \frac{\left\|F^{[r]}-\widetilde F^{[r]}\right\|_F}
 {\left\|F^{[r]}\right\|_F}\le\eps\right\}.
 \label{eq:o1_eps_rank}
\end{equation}
Thus, $D_r^{(\eps)}(F)$ is the smallest bond dimension required across bond $r$ to approximate $F$ with relative Frobenius error $\eps$. Equivalently, it is the approximate Schmidt rank at that cut~\cite{Oseledets2011}.

\subsection{The scalar damping kernel} \label{sec:dampingkernel}

The scalar damping kernel is
\begin{equation}
 \mathcal{D}_N(k,j)=\exp\!\left(-\frac{\omega_rkj}{N}\right),
 \qquad 0\le k,j<N.
 \label{eq:o1_kernel}
\end{equation}
Consider the MPO bond after the first $r$ sites and set $X=2^r$, $Y=2^{n-r}$, so $XY=N$. In the no-SWAP site ordering,
\begin{equation}
 k=k_L+Xk_R,\qquad j=Yj_L+j_R,
 \label{eq:o1_split}
\end{equation}
where $0\le k_L,j_L<X$ and $0\le k_R,j_R<Y$. Here, $k_L$ contains the $r$ least significant bits of $k$, while $j_L$ contains the $r$ most significant bits of $j$. The resulting matrix has row index $(k_L,j_L)$ and column index $(k_R,j_R)$. Direct expansion of the exponent gives
\begin{align}
 \frac{\omega_rkj}{N}
 &=\frac{\omega_rk_Lj_L}{X}
  +\frac{\omega_rk_Rj_R}{Y}
  +\frac{\omega_rk_Lj_R}{N}
  +\omega_rk_Rj_L .
 \label{eq:o1_four_factor}
\end{align}
The first two terms are local to the left and right blocks and, therefore, do not increase its rank.  The coupling across the bond is contained in the two kernels 
\begin{align}
 P_{X,Y}(u,v)&=\e^{-\omega_ruv/N},\nonumber\\
 Q_{X,Y}(u,v)&=\e^{-\omega_ruv},
 \qquad 0\le u<X,\quad 0\le v<Y.
 \label{eq:o1_cross_kernels}
\end{align}
Here $u$ and $v$ are row and column indices. These kernels have different structures. The exponent of $P_{X,Y}$ stays bounded by $\omega_r$, so we use a polynomial approximation. In contrast, $Q_{X,Y}$ decays exponentially with $uv$, and can be approximated by keeping only its leading rows. 

\paragraph*{Polynomial approximation of $P_{X,Y}$.}
For $\omega_r>0$, set
\begin{equation}
 z:=\frac{\omega_r uv}{N},
 \qquad
 b:=\frac{\omega_r}{2},
 \qquad
 s:=\frac{2z}{\omega_r}-1.
 \label{eq:o1_cheb_variables}
\end{equation}
Since $0\le uv<N$, we have $0\le z<\omega_r$ and
$s\in[-1,1)$. Moreover, $z=b(1+s)$, so $P_{X,Y}(u,v) =\e^{-z}=\e^{-b}\e^{-bs}$. Expanding $\e^{-bs}$ in Chebyshev polynomials gives
\begin{equation}
 P_{X,Y}(u,v)
 =\e^{-b}
 \left[
 I_0(b)
 +2\sum_{q=1}^{\infty}
 (-1)^q I_q(b)T_q(s)
 \right],
 \label{eq:o1_cheb_series}
\end{equation}
Here, $T_q$ is the Chebyshev polynomial defined by
$T_q(\cos\vartheta)=\cos(q\vartheta)$, while $I_q$ is the modified Bessel function of the first kind,
\begin{equation}
I_q(b):=
\sum_{p=0}^{\infty}
\frac{(b/2)^{2p+q}}
{p!\,(p+q)!},
\qquad q=0,1,2,\ldots.
\label{eq:o1_bessel_definition}
\end{equation}
Equation~\eqref{eq:o1_cheb_series} follows from the standard Fourier--Bessel expansion of $\exp(-b\cos\vartheta)$ after setting $s=\cos\vartheta$.



Retaining the terms $q=0,\ldots,d-1$ gives a polynomial of degree $d-1$. Since $|T_q(s)|\le1$ for $s\in[-1,1]$, the absolute truncation error is bounded by $2\e^{-b}\sum_{q=d}^{\infty}I_q(b)$.
Dividing by $P_{X,Y}(u,v)=\e^{-z}\ge\e^{-2b}$ gives, for every $0\le u<X$ and $0\le v<Y$, the entrywise relative-error bound
\begin{align}
 |P_{X,Y}(u,v)-\widetilde P^{(d)}(u,v)|
 &\le\delta_d P_{X,Y}(u,v),\nonumber\\
 \delta_d&:=2\e^b\sum_{q=d}^{\infty}I_q(b).
 \label{eq:o1_delta}
\end{align}
The truncated series is a polynomial of degree $d-1$ in $z=(\omega_r/N)uv$. Each monomial factorizes into a function of $u$ and a function of $v$. To make this separation explicit, write the truncated polynomial as $\sum_{\ell=0}^{d-1}c_\ell z^\ell$.  For each $\ell$, define the vectors
\begin{equation}
 \boldsymbol p_\ell:=\bigl(u^\ell\bigr)_{u=0}^{X-1}\in\mathbb R^X,
 \quad
 \boldsymbol q_\ell:=c_\ell\left(\frac{\omega_r}{N}\right)^\ell
 \bigl(v^\ell\bigr)_{v=0}^{Y-1}\in\mathbb R^Y.
 \label{eq:o1_P_vectors}
\end{equation}
Here $u^0=v^0=1$, including when $u=0$ or $v=0$. For a single entry $(u,v)$, $\widetilde P^{(d)}(u,v)
 =\sum_{\ell=0}^{d-1} c_\ell\left(\frac{\omega_r}{N}\right)^\ell  u^\ell v^\ell$.
Hence,
\begin{equation}
 \widetilde P^{(d)}
 =\sum_{\ell=0}^{d-1}\boldsymbol p_\ell\boldsymbol q_\ell^{\mathsf T}.
 \label{eq:o1_P_outer_products}
\end{equation}
Thus, the truncated kernel is a sum of at most $d$ separable terms across the bond. Each term is a rank-one outer product, and therefore
\begin{equation}
 \operatorname{rank}\widetilde P^{(d)}\le d.
 \label{eq:o1_P_rank}
\end{equation}
Both the entrywise relative-error bound $\delta_d$ and the rank bound $d$ are independent of $N$ and the bond position.

\vspace{0.1cm}
\paragraph*{Row-truncation approximation of $Q_{X,Y}$.}
For $v\geq1$, the entries of $Q_{X,Y}$ decay exponentially with the row index $u$. We therefore retain the first $U+1$ rows and keep only the exact $v=0$ entry in the remaining rows. For an integer $U\ge0$, define
\begin{equation}
 \widetilde Q^{(U)}(u,v)=
 \begin{cases}
  \e^{-\omega_ruv},&u\le U,\\
  \mathbf 1_{\{v=0\}},&u>U.
 \end{cases}
 \label{eq:o1_Q_tilde}
\end{equation}
There are $U+1$ retained rows, and all remaining rows are identical. Therefore,
\begin{equation}
 \operatorname{rank}\widetilde Q^{(U)}\le\min\{X,U+2\}.
 \label{eq:o1_Q_rank}
\end{equation}
Only entries with $u>U$ and $v\ge1$ contribute to the error. Extending these sums to infinity gives
\begin{align}
 \|Q_{X,Y}-\widetilde Q^{(U)}\|_F^2
 &\le\sum_{u>U}\sum_{v\ge1}\e^{-2\omega_ruv}\nonumber\\
 &=\sum_{u>U}\frac{\e^{-2\omega_ru}}{1-\e^{-2\omega_ru}}\nonumber\\
 &\le\frac{1}{1-\e^{-2\omega_r(U+1)}}
 \sum_{u>U}\e^{-2\omega_ru}\nonumber\\
 &\le
 \frac{\e^{-2\omega_r(U+1)}}
 {(1-\e^{-2\omega_r})(1-\e^{-2\omega_r(U+1)})}
 =:\eta_U^2 .
 \label{eq:o1_eta}
\end{align}
Thus, for fixed $\omega_r>0$, the Frobenius error can be reduced to any target value by choosing $U$ independently of $N$ and the bond position.

\paragraph*{Combined rank and error bound.}
Let $A=\e^{-\omega_rk_Lj_L/X}$ and
$B=\e^{-\omega_rk_Rj_R/Y}$ be the local factors in
Eq.~\eqref{eq:o1_four_factor}, and define
\begin{equation}
 \widetilde {\mathcal{D}}_N^{[r]}:=A B\,
 \widetilde P^{(d)}(k_L,j_R)\,
 \widetilde Q^{(U)}(j_L,k_R).
 \label{eq:o1_D_tilde}
\end{equation}
Expanding $\widetilde P^{(d)}\widetilde Q^{(U)}$ gives at most $d(U+2)$ separable terms across the bond. Each term is a rank-one outer product in the unfolding. The one-sided factors $A$ and $B$ do not change the rank. Thus,
\begin{equation}
 \operatorname{rank}\widetilde {\mathcal{D}}_N^{[r]}\le d(U+2).
 \label{eq:o1_composed_rank}
\end{equation}
We separate the two errors using
\[
 ABPQ-AB\widetilde P\widetilde Q
 =AB(P-\widetilde P)Q+AB\widetilde P(Q-\widetilde Q).
\]
For every matrix entry, Eq.~\eqref{eq:o1_delta} gives $|AB(P-\widetilde P)Q|\le\delta_d |ABPQ|=\delta_d |\mathcal{D}_N|$; hence approximating $P_{X,Y}$ contributes at most $\delta_d\|\mathcal{D}_N\|_F$.  For the $Q_{X,Y}$ term, $0<A,B\le1$ and $|\widetilde P^{(d)}(k_L,j_R)|\le1+\delta_d$.  Since $Q_{X,Y}-\widetilde Q^{(U)}$ depends only on $(j_L,k_R)$, summing its squared error over the remaining pair $(k_L,j_R)$ gives the multiplicity $XY=N$:
\[
 \|AB\widetilde P(Q-\widetilde Q)\|_F^2
 \le (1+\delta_d)^2N\eta_U^2.
\]
Hence
\begin{equation}
 \|\mathcal{D}_N^{[r]}-\widetilde {\mathcal{D}}_N^{[r]}\|_F
 \le\delta_d\|\mathcal{D}_N\|_F+(1+\delta_d)\sqrt N\,\eta_U.
 \label{eq:o1_composed_error}
\end{equation}
The column $j=0$ contains $N$ ones, so $\|\mathcal{D}_N\|_F\ge\sqrt N$. Since unfolding preserves the Frobenius norm, dividing Eq.~\eqref{eq:o1_composed_error} by $\|\mathcal{D}_N^{[r]}\|_F$ gives the relative-error bound $\delta_d+(1+\delta_d)\eta_U$.



Therefore, if integers $d\ge1$ and $U\ge0$ satisfy
\begin{equation}
 \delta_d+(1+\delta_d)\eta_U\le\eps,
 \label{eq:o1_admissible}
\end{equation}
then, for any internal MPO bond $r$,
\[
 D_r^{(\eps)}(\mathcal{D}_N)\le d(U+2).
\]

Minimizing over all such choices gives
\begin{equation}
 \mathcal B(\omega_r,\eps):=
 \min_{\substack{d\in\mathbb N,\ U\in\mathbb N_0\\
 \delta_d(\omega_r)+[1+\delta_d(\omega_r)]
 \eta_U(\omega_r)\le\eps}}
 d(U+2).
 \label{eq:o1_B}
\end{equation}
For fixed $\omega_r>0$ and $\eps>0$, both $\delta_d$ and $\eta_U$ vanish as $d$ and $U$ increase. Hence, finite values of $d$ and $U$ can always be chosen to satisfy Eq.~\eqref{eq:o1_admissible}, and
\[
 D_r^{(\eps)}(\mathcal{D}_N)\le\mathcal B(\omega_r,\eps).
\]
 Since this bound depends only on $\omega_r$ and $\eps$, not on $N$ or the bond position, the approximate Schmidt rank at fixed accuracy is $O(1)$ in $N$. 
 The numerical bounds below are obtained by evaluating $\delta_d$ and $\eta_U$ directly. For $\omega_r=0$, $\mathcal{D}_N(k,j)=1$, so every bond has exact Schmidt rank one.

\subsection{Accuracy dependence at fixed $\omega_r$} \label{sec:o1_fixed_omega_accuracy}
We next determine how the truncation parameters $d$ and $U$ depend on the accuracy. We fix $\omega_r>0$ and assume $0<\eps\le 1$. For integers $p,q\ge0$, the factorial in Eq.~\eqref{eq:o1_bessel_definition} satisfies
\begin{equation}
 (p+q)!
 =q!\prod_{\ell=1}^{p}(q+\ell)
 \ge q!(q+1)^p.
 \label{eq:o1_shifted_factorial_bound}
\end{equation}
Substitution into Eq.~\eqref{eq:o1_bessel_definition} gives
\begin{align}
 I_q(b)
 &\le \frac{(b/2)^q}{q!}
 \sum_{p=0}^{\infty}
 \frac{1}{p!}\left(\frac{b^2}{4(q+1)}\right)^p\nonumber\\
 &=\frac{(b/2)^q}{q!}
 \exp\!\left(\frac{b^2}{4(q+1)}\right).
 \label{eq:o1_bessel_factorial_bound}
\end{align}
Let
\begin{equation}
 t_q:=\frac{(b/2)^q}{q!},
 \qquad
 \rho_d:=\frac{b}{2(d+1)}.
 \label{eq:o1_tail_parameters}
\end{equation}
For every integer $k\ge0$,
\begin{equation}
 \frac{t_{d+k}}{t_d}
 =\prod_{\ell=1}^{k}\frac{b}{2(d+\ell)}
 \le\rho_d^k.
 \label{eq:o1_tail_ratio_bound}
\end{equation}
We choose $d \geq \e b$. This choice is sufficient for an upper bound, although it is not necessarily optimal. It then follows that
\begin{equation}
 \rho_d<\frac{1}{2\e}<1,
 \qquad
 \exp\!\left(\frac{b^2}{4(d+1)}\right)
 \le\exp\!\left(\frac{b}{4\e}\right).
 \label{eq:o1_m_large_consequences}
\end{equation}
For $q\ge d$,
$\exp[b^2/(4(q+1))]\le\exp[b^2/(4(d+1))]$. Together with Eq.~\eqref{eq:o1_tail_ratio_bound}, this gives
\begin{align}
 \sum_{q=d}^{\infty}I_q(b)
 &\le \exp\!\left(\frac{b^2}{4(d+1)}\right)
 \sum_{q=d}^{\infty}t_q\nonumber\\
 &\le \exp\!\left(\frac{b^2}{4(d+1)}\right)
 t_d\sum_{k=0}^{\infty}\rho_d^k\nonumber\\
 &\le \frac{\exp[b/(4\e)]}{1-1/(2\e)}
 \frac{(b/2)^d}{d!}.
 \label{eq:o1_bessel_tail_bound}
\end{align}
Combining this estimate with Eq.~\eqref{eq:o1_delta} and using
$d!\ge(d/\e)^d$ gives
\begin{align}
 \delta_d
 &\le C_{P,\omega_r}
 \left(\frac{\e b}{2d}\right)^d\nonumber\\
 &\le C_{P,\omega_r}2^{-d},
 \label{eq:o1_delta_stirling_bound}
\end{align}
where
\begin{equation}
 C_{P,\omega_r}
 :=\frac{2\e^{b+b/(4\e)}}{1-1/(2\e)}.
 \label{eq:o1_P_constant}
\end{equation}
We therefore choose
\begin{equation}
 d_\eps
 :=\left\lceil
 \max\left\{\e b,
 \log_2\!\left(\frac{3C_{P,\omega_r}}{\eps}\right)
 \right\}\right\rceil
 \label{eq:o1_m_explicit_choice}
\end{equation}
which ensures $\delta_{d_\eps}\le\eps/3$.

The error in $Q_{X,Y}$ decays exponentially with $U$. Using Eq.~\eqref{eq:o1_eta} and $1-\e^{-2\omega_r(U+1)}\ge 1-\e^{-2\omega_r}$, we obtain
\begin{align}
 \eta_U
 &\le \frac{\e^{-\omega_r(U+1)}}{1-\e^{-2\omega_r}}
 =C_{Q,\omega_r}\e^{-\omega_r U},\nonumber\\
 C_{Q,\omega_r}
 &:=\frac{\e^{-\omega_r}}{1-\e^{-2\omega_r}}.
 \label{eq:o1_eta_exponential_bound}
\end{align}
We thus choose
\begin{equation}
 U_\eps
 :=\left\lceil
 \max\left\{0,
 \frac{1}{\omega_r}
 \log\!\left(\frac{3C_{Q,\omega_r}}{\eps}\right)
 \right\}\right\rceil
 \label{eq:o1_U_explicit_choice}
\end{equation}
which ensures $\eta_{U_\eps}\le\eps/3$.
With these choices, the total error satisfies
\begin{align}
 \delta_{d_\eps}+
 (1+\delta_{d_\eps})\eta_{U_\eps}
 &\le \frac{\eps}{3}
 +\left(1+\frac{\eps}{3}\right)\frac{\eps}{3}\nonumber\\
 &=\frac{2\eps}{3}+\frac{\eps^2}{9}<\eps.
 \label{eq:o1_accuracy_split}
\end{align}
Hence, we obtain the logarithmic accuracy dependence,
\begin{align}
 d_\eps =& \;O_{\omega_r}\!\left(1+\log\frac{1}{\eps}\right), \nonumber \\
 U_\eps =& \; O_{\omega_r}\!\left(1+\log\frac{1}{\eps}\right).
 \label{eq:o1_mU_log_scaling}
\end{align}

\paragraph*{TT--SVD bound for the scalar kernel.} The estimate above is for a single bipartition. To construct a TT-SVD MPO approximation to the scalar kernel with overall error tolerance $\eps$, use the per-bond tolerance
\begin{equation}
 \eps_{\rm bond}:=\frac{\eps}{\sqrt{n-1}}
 \label{eq:o1_kernel_cut_tolerance}
\end{equation}
at each of its $n-1$ internal bonds. By the Eckart--Young theorem~\cite{EckartYoung1936}, the cut-wise rank bound limits the discarded singular-value tail at every bond to $\eps_{\rm bond}\|\mathcal{D}_N\|_F$. The standard TT-SVD error estimate~\cite{Oseledets2011} then gives
\begin{align}
 \frac{\|\mathcal{D}_N-\widetilde {\mathcal{D}}_N\|_F^2}{\|\mathcal{D}_N\|_F^2}
 &\le \sum_{r=1}^{n-1}\eps_{\rm bond}^2\nonumber\\
 &=(n-1)\eps_{\rm bond}^2=\eps^2.
 \label{eq:o1_kernel_global_error}
\end{align}
Using $\eps_{\rm bond}$ in Eq.~\eqref{eq:o1_B} therefore bounds the largest bond dimension of this TT-SVD MPO
\begin{align}
 D_{\mathrm{TT\text{-}SVD}}^{(\eps)}(\mathcal{D}_N)
 &\le\mathcal B\!\left(
  \omega_r,\frac{\eps}{\sqrt{n-1}}
  \right)\nonumber\\
 &=O_{\omega_r}\!\left[
  \left(
   1+\log\frac{1}{\eps}+\frac12\log(n-1)
  \right)^2
  \right]\nonumber\\
 &=O_{\omega_r}\!\left[
  \left(
   1+\log\frac{1}{\eps}+\log\log N
  \right)^2
  \right].
 \label{eq:o1_kernel_global_bond_scaling}
\end{align}
Since $n=\log_2N$, the largest bond dimension of the scalar-kernel TT-SVD MPO scales, at fixed $\omega_r$ and fixed overall error, as $O_{\omega_r,\eps}\left((\log\log N)^2\right)$. This result concerns only the scalar kernel, which has $n-1$ internal bonds. We next extend the cut-wise argument to the two-register $\DT$ operator, whose MPO has $2n$ qubit sites and hence $2n-1$ internal bonds. The site ordering determines the bipartition associated with each bond.

\subsection{Bond-dimension bound for the two-register damping operator}
\label{sec:two_register_bound}

We now consider the two-register operator defined by the no-SWAP gate-network core. The numerical MPO approximates this operator through repeated zip-up/zip-down compressions. Let $h,h'$ denote the input values of the first and second registers, and let $k,k'$ denote their output values, with all four indices in $0,\ldots,N-1$. The corresponding matrix elements are
\begin{equation}
 \langle k,k'|\widehat{\mathrm{DT}}|h,h'\rangle
 =\frac{1}{\sqrt N} \mathcal{D}_N(k;h,h')\,\mathbf 1_{\{k'=h'\}}.
 \label{eq:o1_full_matrix_element}
\end{equation}
The input MPS encodes the time-series data in the copy subspace $h'=h$, but the $\DT$ MPO is defined on the full two-register space. Thus, $h$ and $h'$ are independent in the operator Schmidt-rank analysis. The unchanged second register gives the local constraint
\[
 \mathbf 1_{\{k'=h'\}}
 =\prod_{i=1}^{n}\mathbf 1_{\{k'_i=h'_i\}}.
\]
This product has operator Schmidt rank one across every MPO bond.

\paragraph*{Gate-product form.}
Write $h=(h_1,\ldots,h_n)$ and $h'=(h'_1,\ldots,h'_n)$ in most-significant-bit order. We label the target position by $t$ and the control position by $c$. The control bit at position $c$ has weight $2^{n-c}$, while the target output bit $k_t$ has weight $2^{t-1}$ in the no-SWAP ordering.
For a fixed target $t$, the controlled-damping gates use $h'_c$ when $c<t$ and $h_c$ when $c>t$, while the damping-Hadamard contributes the $c=t$ term. Thus, the binary weights entering target $t$ are drawn from both registers. We collect them in the target-dependent control index
\begin{equation}
 J^{(t)}
 =\sum_{c=1}^{t-1}h'_c2^{n-c} + \sum_{c=t}^{n}h_c2^{n-c}.
 \label{eq:o1_mixed_control_integer}
\end{equation}
Thus, target $t$ contributes $k_t2^{t-1}J^{(t)}/N$ to the exponent. Expanding $J^{(t)}$ over its control bits, the coefficient of each target--control pair is
\[
 \frac{2^{t-1}2^{n-c}}{N}=2^{t-c-1}.
\]
Summing over all targets gives the circuit coefficient
\begin{equation}
 \begin{aligned}
 \mathcal{D}_N(k;h,h')
 =\exp\!\Bigg[-\omega_r\Bigg(&
 \sum_{1\le t\le c\le n}2^{t-c-1}k_t h_c\\[-1mm]
 &+\sum_{1\le c<t\le n}2^{t-c-1}k_t h'_c
 \Bigg)\Bigg].
 \end{aligned}
 \label{eq:o1_circuit_coefficient}
\end{equation}
When $h'=h$, the two sums combine:
\[
 \sum_{t,c=1}^{n}2^{t-c-1}k_t h_c
 =\frac{1}{N}
 \left(\sum_t k_t2^{t-1}\right)
 \left(\sum_c h_c2^{n-c}\right)
 =\frac{kh}{N}.
\]
Hence $\mathcal{D}_N(k;h,h)=\mathcal{D}_N(k,h)$. For $h'\neq h$, the circuit coefficient generally is not a single scalar kernel.

\paragraph*{MPO bond between register pairs.}
The MPO sites are ordered as
$(h_1,k_1),(h'_1,k'_1),\ldots,(h_n,k_n),(h'_n,k'_n)$.
Consider the MPO bond after the first $r$ register pairs,  and set $X=2^r$ and $Y=2^{n-r}$. Among the four indices, only $k$ is represented in reversed bit order:
\begin{equation}
 \begin{aligned}
  k&=k_L+Xk_R, & h&=Yh_L+h_R,\\
  h'&=Yh'_L+h'_R, & k'&=Yk'_L+k'_R.
 \end{aligned}
 \label{eq:o1_four_index_split}
\end{equation}
Thus, $k_L$ contains the low-order bits of $k$, while $h_L$, $h'_L$, and $k'_L$ contain high-order bits of their respective indices. Every term in Eq.~\eqref{eq:o1_circuit_coefficient} is either local to one side of the bond or couples the two sides:
\begin{center}
\begin{tabular}{c|c|c}
 & $c\le r$ & $c>r$\\ \hline
$t\le r$ & left only & crossing, with control $h_c$\\
$t>r$ & crossing, with control $h'_c$ & right only
\end{tabular}
\end{center}
In one crossing block, the control bits come from $h$, while in the other they come from $h'$. Since $2^{t-c-1}=2^{t-1}2^{-c}$, both crossing sums factorize across the bond:
\begin{align*}
 \sum_{t\le r<c}2^{t-c-1}k_t h_c
 &=\frac{1}{N}
 \left(\sum_{t\le r}k_t2^{t-1}\right)
 \left(\sum_{c>r}h_c2^{n-c}\right)\\
 &=\frac{k_Lh_R}{N},\\
 \sum_{c\le r<t}2^{t-c-1}k_t h'_c
 &=\left(\sum_{t>r}k_t2^{t-r-1}\right)
 \left(\sum_{c\le r}h'_c2^{r-c}\right)\\
 &=k_Rh'_L.
\end{align*}
All other terms depend on only one side of the bond. Hence
\begin{equation}
 \mathcal{D}_N(k;h,h')
 =\mathsf L_r\mathsf R_r
 P_{X,Y}(k_L,h_R)Q_{X,Y}(h'_L,k_R).
 \label{eq:o1_circuit_factorization}
\end{equation}
Here, $\mathsf L_r$ and $\mathsf R_r$ depend only on the left and right variables, respectively. Their exponents are non-positive, so $0<\mathsf L_r,\mathsf R_r\le1$. The two factors that couple the two sides of the bond are the same kernels $P_{X,Y}$ and $Q_{X,Y}$ introduced for the scalar kernel, with $j_R=h_R$ and $j_L=h'_L$.
The bond-dimension bound is therefore governed by the approximate ranks of these kernels, rather than directly by the number of controlled-damping gates crossing the bond.

\paragraph*{Rank and error at one MPO bond.}
We replace $P_{X,Y}$ and $Q_{X,Y}$ by the approximations $\widetilde P^{(d)}$ and $\widetilde Q^{(U)}$. Their product has rank at most $d(U+2)$, while the one-sided factors and the local constraint $k'=h'$ do not increase this rank. 

As in the scalar case, the error from $\widetilde P^{(d)}$ is at most $\delta_d\|\mathcal{D}_N\|_F$. The error in $\widetilde Q^{(U)}$ depends only on $(h'_L,k_R)$. Summing its squared error over the remaining indices $(k_L,h_L,h_R,h'_R)$ gives a multiplicity $X^2Y^2=N^2$. Thus, its Frobenius error is at most $(1+\delta_d)N\eta_U$, and
\[
 \|\mathcal{D}_N^{[r]}-\widetilde{\mathcal D}_N^{[r]}\|_F
 \le\delta_d\|\mathcal{D}_N\|_F+(1+\delta_d)N\eta_U.
\]

Here, $\widetilde{\mathcal D}_N^{[r]}$ is obtained from the factorization in Eq.~\eqref{eq:o1_circuit_factorization}. For $k=0$, $\mathcal{D}_N(0;h,h')=1$ for all $N^2$ input pairs $(h,h')$, and $\|\mathcal{D}_N\|_F\ge N$. Therefore, dividing the preceding error by this norm gives the relative error at most $\delta_d+(1+\delta_d)\eta_U$. At bonds between complete register pairs, the constraint $k'=h'$ changes neither the Schmidt rank nor the relative Frobenius error. Therefore, $\widehat{\mathrm{DT}}$ has the same relative error as its coefficient tensor. Hence, any pair $(d,U)$ satisfying Eq.~\eqref{eq:o1_admissible} gives
\begin{equation}
 D_r^{(\eps)}\!\left(\widehat{\mathrm{DT}}\right)
 \le\mathcal B(\omega_r,\eps).
 \label{eq:o1_circuit_bound}
\end{equation}
This bound applies at MPO bonds between complete register pairs. A bond inside register pair $s$ lies between the sites $(h_s,k_s)$ and $(h'_s,k'_s)$. Relative to the bond immediately after pair $s$, or to the right boundary if $s=n$, the site $(h'_s,k'_s)$ lies on the opposite side of the bipartition. The local constraint $k'_s=h'_s$ allows only the two configurations $(0,0)$ and $(1,1)$. Therefore, moving this site can increase the rank by at most a factor of two. This reshaping does not change the Frobenius error. Hence every internal bond of the $\DT$ operator in the no-SWAP site ordering satisfies
\begin{equation}
 D_r^{(\eps)}\!\left(\widehat{\mathrm{DT}}\right)
 \le2\mathcal B(\omega_r,\eps).
 \label{eq:o1_all_circuit_cuts}
\end{equation}
The factor of two is a worst-case bound. In our numerical MPOs, the inside-pair bond dimension is generally not larger than the bond dimension after the complete pairs. We do not analyze this difference further.

\subsection{Overall MPO error and bond-dimension bounds}
\label{sec:complete_mpo_accuracy}

The preceding bounds apply to one MPO bond. To construct an MPO with relative Frobenius error $\eps$, use
\[
 \eps_{\rm bond}=\frac{\eps}{\sqrt{2n-1}}
\]
at each of the $2n-1$ internal cuts. By Eckart--Young theorem~\cite{EckartYoung1936}, rank $2\mathcal B(\omega_r,\eps_{\rm bond})$ gives a discarded singular-value tail of at most $\eps_{\rm bond}\|\widehat{\mathrm{DT}}\|_F$ at every cut. TT-SVD with these ranks produces an MPO whose bond dimensions do not exceed them. Its error estimate~\cite{Oseledets2011} gives the following
\[
 \left\|\widehat{\mathrm{DT}}-\widetilde{\mathrm{DT}}\right\|_F^2
 \le (2n-1)\eps_{\rm bond}^2
 \left\|\widehat{\mathrm{DT}}\right\|_F^2
 =\eps^2\left\|\widehat{\mathrm{DT}}\right\|_F^2.
\]
Let $D_{\mathrm{TT\text{-}SVD}}^{(\eps)}$ denote the largest internal bond dimension of the TT-SVD MPO with overall error tolerance $\eps$.  Then
\begin{equation}
 D_{\mathrm{TT\text{-}SVD}}^{(\eps)} \!\left(\widehat{\mathrm{DT}}\right)
 \le2\mathcal B\!\left(\omega_r,\frac{\eps}{\sqrt{2n-1}}\right).
 \label{eq:o1_global_bound}
\end{equation}

Using Eq.~\eqref{eq:o1_mU_log_scaling} in this bound gives
\begin{align}
 D_{\mathrm{TT\text{-}SVD}}^{(\eps)} \!\left(\widehat{\mathrm{DT}}\right)
 &=O_{\omega_r}\!\left[
  \left(
   1+\log\frac{1}{\eps}+\frac12\log(2n-1)
  \right)^2
  \right]\nonumber\\
 &=O_{\omega_r}\!\left[
  \left(
   1+\log\frac{1}{\eps}+\log\log N
  \right)^2
  \right],
 \label{eq:o1_global_loglog_scaling}
\end{align}
where $n=\log_2N$. In particular, at fixed $\omega_r>0$ and fixed overall error tolerance,
\begin{equation}
 D_{\mathrm{TT\text{-}SVD}}^{(\eps)} \!\left(\widehat{\mathrm{DT}}\right)
 =O_{\omega_r,\eps}\!\left((\log\log N)^2\right).
 \label{eq:o1_global_fixed_accuracy_scaling}
\end{equation}

For the no-SWAP QFT core, the analogous $Q$-type crossing factor is $\exp(-2\pi\im\,\ell_Rj_L)=1$ because $\ell_Rj_L$ is an integer. The remaining $P$-type factor has bounded phase and admits the same polynomial approximation as above, now with ordinary Bessel coefficients; using the per-bond tolerance therefore gives $D_{\mathrm{QFT}}^{(\eps)}=O_{\eps}(\log\log N)$.

\paragraph*{Numerical evaluation at the given parameters.}
The implementation uses the relative SVD truncation threshold $\tau=10^{-15}$. At each truncation, $\tau$ bounds the discarded squared singular values relative to the total squared singular value weight: ${\sum_{q>D}\sigma_q^2}/{\sum_q\sigma_q^2}\le\tau $.
The corresponding per-truncation relative Frobenius error tolerance is $\eps_{\rm SVD}:=\sqrt\tau\simeq3.1623\times10^{-8}$. This is a local scale, not a bound on the final error after several compression sweeps. Let
$\mathcal E_{d,U}:=\delta_d+(1+\delta_d)\eta_U$.
At $\omega_r=2\pi$, direct evaluation gives
\begin{equation}
\begin{alignedat}{2}
 \delta_{15}&\simeq3.9953\times10^{-8},\quad&
 \delta_{16}&\simeq3.8637\times10^{-9},\\
 \eta_1&\simeq3.4873\times10^{-6},&
 \eta_2&\simeq6.5124\times10^{-9},\\
 \eta_3&\simeq1.2162\times10^{-11},&&\\
 \mathcal E_{16,2}&\simeq1.0376\times10^{-8},&
 \mathcal E_{16,3}&\simeq3.8759\times10^{-9}.
\end{alignedat}
\label{eq:o1_B_local_values}
\end{equation}
Since $\delta_d$ and $\eta_U$ are nonnegative and decrease with $d$ and $U$, the inequalities $\delta_{15}>\eps_{\rm SVD}$ and $\eta_1>\eps_{\rm SVD}$ require $d\ge16$ and $U\ge2$. Moreover, $\mathcal E_{16,2}<\eps_{\rm SVD}$, so $(d,U)= (16,2)$ meets the required error tolerance. Therefore, $\mathcal B(2\pi,\eps_{\rm SVD})=16(2+2)=64$, and the bond-dimension bound at every bond is $128$.

For comparison, we choose the overall TT-SVD error tolerance $\eps$ to have the same numerical value as the per-truncation scale, $\eps=\eps_{\rm SVD}$. For $n=30$, the corresponding per-bond
tolerance is
\begin{equation}
 \eps_{\rm bond}
 =\frac{\eps_{\rm SVD}}{\sqrt{2n-1}}
 =\frac{\eps_{\rm SVD}}{\sqrt{59}}
 \simeq4.1169\times10^{-9}.
\end{equation}
Since $\delta_{15}>\eps_{\rm bond}$ and $\eta_2>\eps_{\rm bond}$, we require $d\ge16$ and $U\ge3$. Moreover, $\mathcal E_{16,3}<\eps_{\rm bond}$, so $\mathcal B(2\pi,\eps_{\rm bond})=16(3+2)=80$.
Equation~\eqref{eq:o1_global_bound} therefore gives
\begin{equation}
 D_{\mathrm{TT\text{-}SVD}}^{(\eps)} \!\left(\widehat{\mathrm{DT}}\right)
 \le 2\mathcal B(2\pi,\eps_{\rm bond})=160.
\end{equation}
The observed maximum bond dimension is $18$. It is obtained from repeated local truncations at cutoff $\tau$ and therefore uses a different error criterion from the TT-SVD bound for the complete MPO above. 

\subsection{$\omega_r$-independent bound}\label{sec:koyama} 

The $N$-independent approximate-rank bound above holds for fixed $\omega_r$, but its prefactors become large as $\omega_r\to0^+$ or for very large $\omega_r$. A complementary bound that is uniform for all $\omega_r\ge0$ is derived below. Figure~\ref{fig:o1_uniform_roadmap} summarizes the main steps of this proof.

\begin{figure}[!t]
\centering
\begin{tikzpicture}[
  node distance=0.38cm,
  box/.style={
    rounded corners,
    draw=dtblue,
    fill=dtlight,
    thick,
    align=center,
    text width=0.82\columnwidth,
    minimum height=0.72cm,
    inner sep=4pt,
    font=\footnotesize
  },
  arr/.style={
    -{Latex[length=1.8mm]},
    thick,
    draw=dtblue
  }
]

\node[box] (cut)
  {Bit cut in the scalar damping kernel\\
   $k=k_L+Xk_R$, \qquad $j=Yj_L+j_R$};

\node[box, below=of cut] (cross)
  {Two crossing kernels\\
   $K_{\omega_r/N}(k_L,j_R)$
   and
   $K_{\omega_r}(j_L,k_R)$};

\node[box, below=of cross] (gram)
  {For $K_\nu(u,v)=\e^{-\nu uv}$, form the Gram matrix\\
   $G_\nu=K_\nu K_\nu^{\mathsf T}$, \qquad
   $G_\nu(u,u')=\Psi_Y\!\left(\nu(u+u')\right)$};

\node[box, below=of gram] (cm)
  {Finite completely monotone function\\
   $\Psi_Y(t)=1+f_Y(t)$, \qquad
   $f_Y(t)=\displaystyle\sum_{v=1}^{Y-1}\e^{-vt}$};

\node[box, below=of cm] (exp)
  {Koyama exponential-sum approximation\\
   $f_Y(t)\approx
   \displaystyle\sum_{\ell=1}^{g}
   w_\ell\e^{-\alpha_\ell t}$};

\node[box, below=of exp] (gramrank)
  {Low-rank approximation of the Gram matrix\\
   $\operatorname{rank}(G_{\nu,g+1})\le g+1$,
   with controlled operator-norm error};

\node[box, below=of gramrank] (sv)
  {Control of the crossing-kernel singular values\\
   min--max principle and Eckart--Young give
   $\operatorname{rank}(\widetilde K_\nu)\le g+1$ and
   $\|K_\nu-\widetilde K_\nu\|_F\le\zeta$};

\node[box, below=of sv] (combine)
  {Combine the two crossing kernels\\
   $\operatorname{rank}\!\left(
   \widetilde K_{\omega_r/N}\otimes
   \widetilde K_{\omega_r}\right)\le (g+1)^2$};

\node[box, below=of combine] (mpo)
  {For any fixed bond $r$, the bond dimension required to achieve tolerance $\eps$ scales, uniformly in $\omega_r$, as  $D_r^{(\eps)}
\le 2(g+1)^2 =O\!\left((\log N)^4\right)$};

\draw[arr] (cut) -- (cross);
\draw[arr] (cross) -- (gram);
\draw[arr] (gram) -- (cm);
\draw[arr] (cm) -- (exp);
\draw[arr] (exp) -- (gramrank);
\draw[arr] (gramrank) -- (sv);
\draw[arr] (sv) -- (combine);
\draw[arr] (combine) -- (mpo);

\end{tikzpicture}
\caption{Roadmap of the bond-dimension bound that is uniform in $\omega_r$. A binary cut produces two finite Laplace kernels. Their Gram matrices have a Hankel form generated by finite completely monotone functions. Exponential-sum approximations give low-rank Gram matrices, which control the singular-value tails of the crossing kernels. Combining the two kernel approximations gives the bond-dimension bound at every bond.}
\label{fig:o1_uniform_roadmap}
\end{figure}

We now prove the bound outlined in Fig.~\ref{fig:o1_uniform_roadmap}. Consider the crossing kernel $K_\nu(u,v)=\e^{-\nu uv}$, with $\nu\in\{\omega_r/N,\omega_r\}$. Its Gram matrix is
\begin{align}
 (K_\nu K_\nu^{\mathsf T})(u,u')
 &=\Psi_Y(\nu(u+u')),\nonumber\\
 \Psi_Y(t)&=1+\sum_{v=1}^{Y-1}\e^{-vt}.
 \label{eq:o1_koyama_gram}
\end{align}
The cases $Y=1$ and $Y=2$ are exact with at most two exponential terms. Hence we assume $Y\ge 3$ below. Let $W_Y$ be the step function with unit jumps at
$1,\ldots,Y-1$. Then
\begin{equation}
 \begin{aligned}
  f_Y(t)
  &:=\Psi_Y(t)-1
  =\int_{1/2}^{Y-1/2}\e^{-\xi t}\,dW_Y(\xi),\\
  f_Y(0)&=Y-1.
 \end{aligned}
\end{equation}
Thus, $f_Y$ is completely monotone and has the Laplace--Stieltjes
form required by Koyama's result. Applying Corollary~3.15 of
Ref.~\cite{Koyama2023} to the interval $[1/2,Y-1/2]$ gives an
$g$-term exponential-sum approximation satisfying
\begin{equation}
 \sup_{t\ge0}
 \left|f_Y(t)-\widetilde f_{Y,g}(t)\right|
 <\frac{16}{\pi}(Y-1)\widehat\rho_Y^{-2g},
\end{equation}
where
\begin{equation}
 \widehat\rho_Y=
 \exp\!\left[
  \pi\frac{K_{\rm ell}((2Y-1)^{-1})}
  {K_{\rm ell}\!\left(\sqrt{1-(2Y-1)^{-2}}\right)}
 \right],
\end{equation}
and $K_{\rm ell}$ denotes the complete elliptic integral of the first kind.

The constant term in $\Psi_Y=1+f_Y$ is kept exactly. To make the
error in each Gram-matrix entry at most $\zeta^2/X^2$, it is
sufficient to choose a positive integer $g$ satisfying
\begin{equation}
 g\ge
 \frac{\log\!\left[16(Y-1)X^2/(\pi\zeta^2)\right]}
 {2\log\widehat\rho_Y}.
\end{equation}
The resulting Gram-matrix approximation has rank at most $g+1$. If $g\ge Y-1$, we instead use the original finite sum, whose rank is
at most $Y\le g+1$. Standard bounds for the elliptic integrals give
\begin{equation}
 \frac{1}{\log\widehat\rho_Y}=O(1+\log Y).
\end{equation}
Taking the smallest integer $g$ satisfying the bound above and using
$X,Y\le N$, we obtain
\begin{equation}
 g+1 \le C(1+\log N)
 \left[1+\log N+\log(1/\zeta)\right].
\end{equation}

Each exponential factorizes into a function of $u$ and a function of $u'$, and therefore gives a rank-one outer product. Hence the approximate Gram matrix $G_{g+1}$ has rank at most $g+1$. The Gram-matrix error has $X^2$ entries, so
\begin{equation}
 \left\|K_\nu K_\nu^{\mathsf T}-G_{g+1}\right\|_2
 \le
 \left\|K_\nu K_\nu^{\mathsf T}-G_{g+1}\right\|_F
 \le\frac{\zeta^2}{X}.
\end{equation}
The Gram eigenvalues are $\sigma_j(K_\nu)^2$. The min--max principle therefore gives $\sigma_j(K_\nu)^2\le\zeta^2/X$ for $j>g+1$; summing over at most $X$ singular values and using Eckart--Young~\cite{EckartYoung1936} produces a rank-${(g+1)}$ matrix $\widetilde K_\nu$ with $\|K_\nu-\widetilde K_\nu\|_F\le\zeta$.

Because the two coupling kernels depend on disjoint left--right index pairs, their product becomes, after a permutation of the right indices, $\mathcal C=K_{\omega_r/N}\otimes K_{\omega_r}$.  Define $\widetilde{\mathcal C} =\widetilde K_{\omega_r/N}\otimes\widetilde K_{\omega_r}$. Then
\[
 \mathcal C-\widetilde{\mathcal C}
 =(K_{\omega_r/N}-\widetilde K_{\omega_r/N})\otimes K_{\omega_r}
 +\widetilde K_{\omega_r/N}\otimes
  (K_{\omega_r}-\widetilde K_{\omega_r}).
\]
Since $\|K_\nu\|_F\le\sqrt N$ and $\|\widetilde K_\nu\|_F\le\sqrt N+\zeta$, this identity gives
\[
 \|\mathcal C-\widetilde{\mathcal C}\|_F
 \le2\zeta\sqrt N+\zeta^2,\qquad
 \operatorname{rank}\widetilde{\mathcal C}\le (g+1)^2.
\]
The left- and right-local damping factors are at most one, while $\|\mathcal{D}_N\|_F\ge\sqrt N$; hence the relative kernel error is at most
$2\zeta+\zeta^2/\sqrt N$.  The no-SWAP circuit-core factorization \eqref{eq:o1_circuit_factorization} contains the same two coupling matrices, and the same Frobenius-norm counting gives this relative estimate for the coefficient tensor of the two-register no-SWAP $\DT$ operator.  Taking $\zeta=\eps/3$ gives $2\zeta+\zeta^2/\sqrt N \le\eps(2/3+\eps/9)<\eps$. Thus, at every cut between complete register pairs,
\begin{equation}
 D_r^{(\eps)}\!\left(\widehat{\mathrm{DT}}\right)
 \le C\left\{(1+\log N)
 [1+\log N+\log(1/\eps)]\right\}^{2},
 \label{eq:o1_koyama_bound}
\end{equation}
and a cut inside a register pair introduces at most the already stated factor of two. At fixed accuracy, each coupling matrix has $g+1=O((\log N)^2)$, and multiplying the two ranks gives $(g+1)^2=O((\log N)^4)$ with a constant independent of $\omega_r$.  Although weaker in its $N$ dependence, this bound applies uniformly to all $\omega_r \ge0$ ; at fixed $\omega_r>0$, Eq.~\eqref{eq:o1_circuit_bound} gives the sharper result.

Applying the same TT-SVD error allocation as above, with $\eps_{\rm bond}=\eps/\sqrt{2n-1}$, gives a TT-SVD MPO with global relative error $\eps$. The additional $\log\log N$ contribution from $\eps_{\rm bond}$ does not change the leading scaling, so its maximum bond dimension is $O_{\eps}((\log N)^4)$ uniformly in $\omega_r\geq0$.

These analytical results bound the approximate Schmidt rank across each bipartition and the bond dimensions of the complete MPO constructed by TT-SVD. They do not analyze error accumulation in sequential zip-up/zip-down compression. 

For $\DT$ and $\QFT$, the low bond-dimension bounds arise because the kernels that describes the combined effect of the gates that cross each cut are compressible at the target accuracy. Whether other non-unitary transforms have similarly compressible cross-cut kernels, or different structures that yield compact MPOs, remains an open question.

\section{Circuit implementation of the damping transform}
For target qubit $\ell$, the controlled-damping gate associated with bit $m\ne\ell$ has strength
\begin{equation}
 \theta_{\ell,m}
 =\omega_r2^{\ell-m-1}
 =\frac{\omega_r}{2^{m-\ell+1}}.
 \label{eq:dt_control_angle}
\end{equation}
For $m>\ell$, the control is qubit $m$ of the first register, which has not yet been changed. For $m<\ell$, the control is the unchanged copy qubit $m'$ in the second register; we denote this gate by $R_{\ell,m'}$. The damping-Hadamard gate $\mathcal H_\ell$ supplies the contribution from $m=\ell$.

\subsection{First Qubit}
\begin{enumerate}
\item Start with the state $\ket{j_1\,j_2\,\dots\,j_n} \ket{{j'}_1\,\dots\,{j'}_n}$.
\item Apply $\mathcal{H}_1$ to qubit 1 ($\ket{j_1}$):
  \begin{equation}
     \ket{j_1} \;\longrightarrow\;
     \frac{1}{\sqrt{2}}
     \Bigl(
       \ket{0} + e^{-\omega_r\,j_1/2}\,\ket{1}
     \Bigr).
  \end{equation}
  This yields
  \[
    \frac{1}{\sqrt{2}}
    \Bigl(
      \ket{0} + e^{-\omega_r\,j_1/2}\,\ket{1}
    \Bigr)
    \;
    \ket{j_2\,j_3\,\dots\,j_n} \; \ket{{j'}_1\,\dots\,{j'}_n}.
  \]
\item Apply $R_{1,m}$ for $m = 2,\dots,n$:
  Each gate accumulates the contribution from the qubit $m$ into the exponent for the first qubit's $\ket{1}$ component.

\begin{align}
&\frac{1}{\sqrt{2}}\!\left(\ket{0}+e^{-\omega_r j_1/2}\ket{1}\right)
 \ket{j_2\cdots j_n} \ket{{j'}_1\,\dots\,{j'}_n}
\xrightarrow{\,R_{12}\,}\nonumber\\
&  \frac{1}{\sqrt{2}}\!\left(\ket{0}+e^{-\omega_r (j_1/2 + j_2/2^{2})}\ket{1}\right)
 \ket{j_2\cdots j_n} \ket{{j'}_1\,\dots\,{j'}_n}.
\end{align}

After applying all these gates, the factor yields
  \[
     \exp\Bigl[
       -\omega_r\,
       \Bigl(
         j_1\,2^{-1} + j_2\,2^{-2} + \dots + j_n\,2^{-n}
       \Bigr)
     \Bigr]
     =
     \exp\bigl(
       -\omega_r\,\frac{j}{2^n}
     \bigr),
  \]
  where $j = j_1\,2^{n-1} + \dots + j_n\,2^0$. Thus, the first qubit state transform to
  \[
    \frac{1}{\sqrt{2}}
    \Bigl(
      \ket{0} + e^{-\omega_r\,j/2^n}\,\ket{1}
    \Bigr),
  \]
  and the combined state becomes
  \[
    \frac{1}{\sqrt{2}}
    \Bigl(
      \ket{0} + e^{-\omega_r\,j/2^n}\,\ket{1}
    \Bigr) \;
    \ket{j_2\,j_3\,\dots\,j_n}  \ket{{j'}_1\,\dots\,{j'}_n}.
  \]
\end{enumerate}

\subsection{Second Qubit}
\begin{enumerate}
\item Next, apply $\mathcal{H}$ to $\ket{j_2}$:
  \[
     \ket{j_2}
     \;\longrightarrow\;
     \frac{1}{\sqrt{2}}
     \Bigl(
       \ket{0} + e^{-\omega_r\,j_2/2}\,\ket{1}
     \Bigr).
  \]
\item Apply $R_{2,m}$ for $m \neq 2$, i.e., $m = 1, 3, 4, \dots, n$. Unlike the QFT, which uses control qubits $m > 2$, the damping transform requires controls from all other qubits, including $m = 1$, due to the real exponent. The second register's copy of the state preserves the original qubit states, ensuring the correct damping exponent despite prior gate operations on qubit 1. After these gates, the second-qubit factor is

  \[
    \frac{1}{\sqrt{2}}
    \Bigl(
      \ket{0} + e^{-\omega_r\,j/2^{n-1}}\,\ket{1}
    \Bigr).
  \]
Thus, the state after transforming the first two qubits is
\begin{align}
    &\frac{1}{\sqrt{2}}
    (\ket{0} + e^{-\omega_r\,j/2^n}\,\ket{1})
  \frac{1}{\sqrt{2}}
    (
      \ket{0} + e^{-\omega_r\,j/2^{n-1}}\,\ket{1}
    ) \nonumber\\
    & \qquad\qquad\qquad\qquad\ket{j_3\,\dots\,j_n} \ket{{j'}_1\,\dots\,{j'}_n}
    \end{align}
\end{enumerate}
\vspace{0.1cm}

\subsection{Remaining Qubits}
Repeat the same procedure for the remaining qubits $\ell = 3, 4, \dots, n$, each time applying the damping Hadamard gate $\mathcal{H}_\ell$ followed by the controlled damping gates $R_{\ell,m}$. The final state becomes
\begin{align}
    \frac{1}{\sqrt{2}}&
    (\ket{0} + e^{-\omega_r\,j/2^n}\,\ket{1})\,
  \frac{1}{\sqrt{2}}
    (
      \ket{0} + e^{-\omega_r\,j/2^{n-1}}\,\ket{1}
    ) \;\cdots \nonumber\\
    &\qquad\frac{1}{\sqrt{2}}
    (
      \ket{0} + e^{-\omega_r\,j/2^1}\,\ket{1}
    )  \ket{{j'}_1\,\dots\,{j'}_n}
    \label{eq_circuit_final}
    \end{align}
    Thus, the circuit above produces the output vector of the $\DT$ transform defined in Eq.~\eqref{eq_z_final}, but with the qubit order reversed, requiring a swap of qubits to match the specified ordering.

\subsection{An Example of $\DT$ for $n=3$}   
For $n =3$, Eq.~\eqref{eq_dt} yields,

\begin{equation}
    \ket{j}\;\ket{j'}
    \;\longmapsto\;
    \frac{1}{\sqrt{2^3}}
    \sum_{k=0}^{2^3-1}
    \exp\Bigl(\frac{-\omega_r\,k\,j}{2^3}\Bigr)\,
    \ket{k}\;\ket{j'}.
\end{equation}
Here, $\ket{j'}$ denotes the second register, an identical copy of $\ket{j}$, which is left unchanged by this step. Write $k$ in binary as $k = k_1\,2^{2} + k_2\,2^{1} + k_3\,2^0$ (where each $k_\ell \in \{0,1\}$). Then

\begin{widetext}
\begin{align} \label{eq_z_final_n3}
    \frac{1}{\sqrt{2^3}}\sum_{k=0}^{2^3-1}
   e^{\frac{-\omega_r\,k\,j}{2^3}}\,
   \ket{k} \ket{j'}
     &= \frac{1}{\sqrt{2^3}}\sum_{k_1=0}^1\sum_{k_2=0}^1  \sum_{k_3=0}^1
   \exp\Bigl(-\omega_r\,j\,(k_1 2^{-1}+ k_2 2^{-2} + k_3 2^{-3})\Bigr)\,
   \ket{k_1\,k_2\,k_3} \;\ket{{j'}_1 {j'}_2 {j'}_3}\nonumber\\
   &=
   \frac{1}{\sqrt{2^3}}\bigl(\ket{0} + e^{-\omega_r\,j/2^{1}}\ket{1}\bigr)
   \;
   \bigl(\ket{0} + e^{-\omega_r\,j/2^{2}}\ket{1}\bigr)
   \;
   \bigl(\ket{0} + e^{-\omega_r\,j/2^{3}}\ket{1}\bigr) \ket{{j'}_1 {j'}_2 {j'}_3}.
\end{align}

We now verify that the DT circuit implements Eq.~\eqref{eq_z_final_n3}.

Qubit 1:

\begin{align*}
\ket{j_1 j_2 j_3}&\ket{{j'}_1 {j'}_2 {j'}_3}
      \overset{\mathcal{H}}{\longrightarrow}
     \frac{1}{\sqrt{2}}
     \Bigl(
       \ket{0} + e^{-\omega_r\,j_1/2}\,\ket{1}
     \Bigr) \ket{j_2 j_3} \ket{{j'}_1 {j'}_2 {j'}_3}\;\overset{\mathcal{R}_{12}}{\longrightarrow} \frac{1}{\sqrt{2}} \left( |0\rangle + e^{-\omega_r (j_1/2 +j_2/2^{2})} \ket{1}\right) \ket{j_2 j_3}\; \ket{{j'}_1 {j'}_2 {j'}_3}\nonumber\\
     &\overset{\mathcal{R}_{13}}{\longrightarrow} \frac{1}{\sqrt{2}} \left( |0\rangle + e^{-\omega_r (j_1/2 +j_2/2^{2} +j_3/2^{3})} \ket{1}\right) \ket{j_2 j_3} \ket{{j'}_1 {j'}_2 {j'}_3} = \frac{1}{\sqrt{2}} \left( |0\rangle + e^{-\omega_r j/2^3} \ket{1}\right) \ket{j_2 j_3}\ket{{j'}_1 {j'}_2 {j'}_3}
  \end{align*}

  Qubit 2:
\begin{align*}
  \frac{1}{\sqrt{2}} \left( |0\rangle + e^{-\omega_r j/2^3} \ket{1}\right)& \ket{j_2 j_3}\; \ket{{j'}_1 {j'}_2 {j'}_3}\;\overset{\mathcal{H}}{\longrightarrow}
     \frac{1}{\sqrt{2}} \left( |0\rangle + e^{-\omega_r j/2^3}\ket{1} \right) \frac{1}{\sqrt{2}} \left( |0\rangle + e^{-\omega_r j_2/2}\ket{1} \right) \ket{ j_3}\; \ket{{j'}_1 {j'}_2 {j'}_3}\nonumber\\
     &\overset{\mathcal{R}_{23}}{\longrightarrow} 
     \frac{1}{\sqrt{2}} \left( |0\rangle + e^{-\omega_r j/2^3}\ket{1} \right) \frac{1}{\sqrt{2}} \left( |0\rangle + e^{-\omega_r (j_2/2+j_3/2^2)}\ket{1} \right) \ket{ j_3}\; \ket{{j'}_1 {j'}_2 {j'}_3}\nonumber\\
     & \overset{\mathcal{R}_{21'}}{\longrightarrow} 
     \frac{1}{\sqrt{2}} \left( |0\rangle + e^{-\omega_r j/2^3}\ket{1} \right) \frac{1}{\sqrt{2}} \left( |0\rangle + e^{-\omega_r (j_2/2+j_3/2^2 + {j'}_1/2^0)}\ket{1} \right) \ket{ j_3}\; \ket{{j'}_1 {j'}_2 {j'}_3}\nonumber\\
     & = 
     \frac{1}{\sqrt{2}} \left( |0\rangle + e^{-\omega_r j/2^3}\ket{1} \right) \frac{1}{\sqrt{2}} \left( |0\rangle + e^{-\omega_r j/2^2}\ket{1} \right) \ket{ j_3} \;\ket{{j'}_1 {j'}_2 {j'}_3} 
  \end{align*}
  
  We use ${j'}_1=j_1$, as ${j'}_1$ is a copy of $j_1$, to obtain the above equation.
  \vspace{1cm}

  Qubit 3:

\begin{eqnarray}
& &  \frac{1}{\sqrt{2}}\left( |0\rangle + e^{-\omega_r j/2^3}\ket{1} \right) \frac{1}{\sqrt{2}} \left( |0\rangle + e^{-\omega_r j/2^2}\ket{1} \right) \ket{ j_3} \;\ket{{j'}_1 {j'}_2 {j'}_3}\nonumber\\
  & & \overset{\mathcal{H}}{\longrightarrow}
    \frac{1}{\sqrt{2}} \left( |0\rangle + e^{-\omega_r j/2^3}\ket{1} \right) \frac{1}{\sqrt{2}} \left( |0\rangle + e^{-\omega_r j/2^2}\ket{1} \right)  \frac{1}{\sqrt{2}}\left( |0\rangle + e^{-\omega_r j_3/2}\ket{1} \right) \;\ket{{j'}_1 {j'}_2 {j'}_3}\nonumber\\
    & &\overset{\mathcal{R}_{31'}}{\longrightarrow}
    \frac{1}{\sqrt{2}} \left( |0\rangle + e^{-\omega_r j/2^3}\ket{1} \right) \frac{1}{\sqrt{2}} \left( |0\rangle + e^{-\omega_r j/2^2}\ket{1} \right)  \frac{1}{\sqrt{2}}\left( |0\rangle + e^{-\omega_r ( j_3/2+ {j'}_1/2^{-1})}\ket{1} \right) \;\ket{{j'}_1 {j'}_2 {j'}_3}\nonumber\\
   & &\overset{\mathcal{R}_{32'}}{\longrightarrow}
    \frac{1}{\sqrt{2}} \left( |0\rangle + e^{-\omega_r j/2^3}\ket{1} \right) \frac{1}{\sqrt{2}} \left( |0\rangle + e^{-\omega_r j/2^2}\ket{1} \right)  \frac{1}{\sqrt{2}}\left( |0\rangle + e^{-\omega_r (j_3/2+{j'}_1/2^{-1}+ {j'}_2/2^{0})}\ket{1} \right) \;\ket{{j'}_1 {j'}_2 {j'}_3}\nonumber\\
    & & = \frac{1}{\sqrt{2}} \left( |0\rangle + e^{-\omega_r j/2^3}\ket{1} \right) \frac{1}{\sqrt{2}} \left( |0\rangle + e^{-\omega_r j/2^2}\ket{1} \right)  \frac{1}{\sqrt{2}}\left( |0\rangle + e^{-\omega_r j/2}\ket{1} \right) \ket{{j'}_1 {j'}_2 {j'}_3}
  \end{eqnarray}
  \end{widetext}
  which matches Eq.~\eqref{eq_z_final_n3} up to a swap of the qubit order. 
  
  The transformation is not unitary, but it is still linear. Since it implements the correct transformation on each element of the computational basis, it also performs it on its superpositions. 

  \section{Input state MPS}\label{app:initialization}
To prepare the input state MPS for the state (Eq.~(3) of the main text),
\[
|x\rangle=\sum_{i} x_i|i\rangle_A|i\rangle_B
= \sum_{i_1,\ldots,i_n} x_{i_1\cdots i_n}|i_1\cdots i_n\rangle_A|i_1\cdots i_n\rangle_B,
\]
we first encode the data as
\[
|\psi\rangle=\sum_{i=0}^{2^n-1} x_i\,|i\rangle
= \sum_{i_1,\ldots,i_n\in\{0,1\}} x_{i_1\cdots i_n}\,|i_1\cdots i_n\rangle,
\]
and, from its MPS, build the MPS for the resulting \(2n\)-qubit state. If this
\(2n\)-qubit state is decomposed into an MPS in the order
\((i_1,\ldots,i_n,i'_1,\ldots,i'_n)\), long-range \(A/B\) correlations drive
exponential growth of the MPS bond dimension, negating any advantage of a
z-transform using a compressed MPO. In contrast, the interleaved order
\((i_1,i'_1,i_2,i'_2,\ldots,i_n,i'_n)\) localizes the copy constraints
\(\delta_{i_k,i'_k}\) and keeps the MPS bond dimension small for
\emph{structured} signals (e.g., a few damped sinusoids), whereas generic random signals the bond dimension remains large. To avoid materializing the \(2^{2n}\)
tensor, we first build an \(n\)-qubit MPS for \(|\psi\rangle\) and then
\emph{lift} it to \(2n\) sites by inserting local copy tensors; this preserves
compression, for example, \(x_j=\sin(\omega_0 j)\) yields bond-dimension, \(\chi=2\) on \(n\) qubits
and bond-dimension, \(\chi\approx 4\) after lifting, making the subsequent $\ZT$ MPO
application tractable.

\section{Constructing the Matrix Product Operator}
\label{app:MPO_building}
\renewcommand{\thefigure}{C\arabic{figure}}
\setcounter{figure}{0}

\begin{figure}[t]
    \centering
    \includegraphics[width=0.98\linewidth]{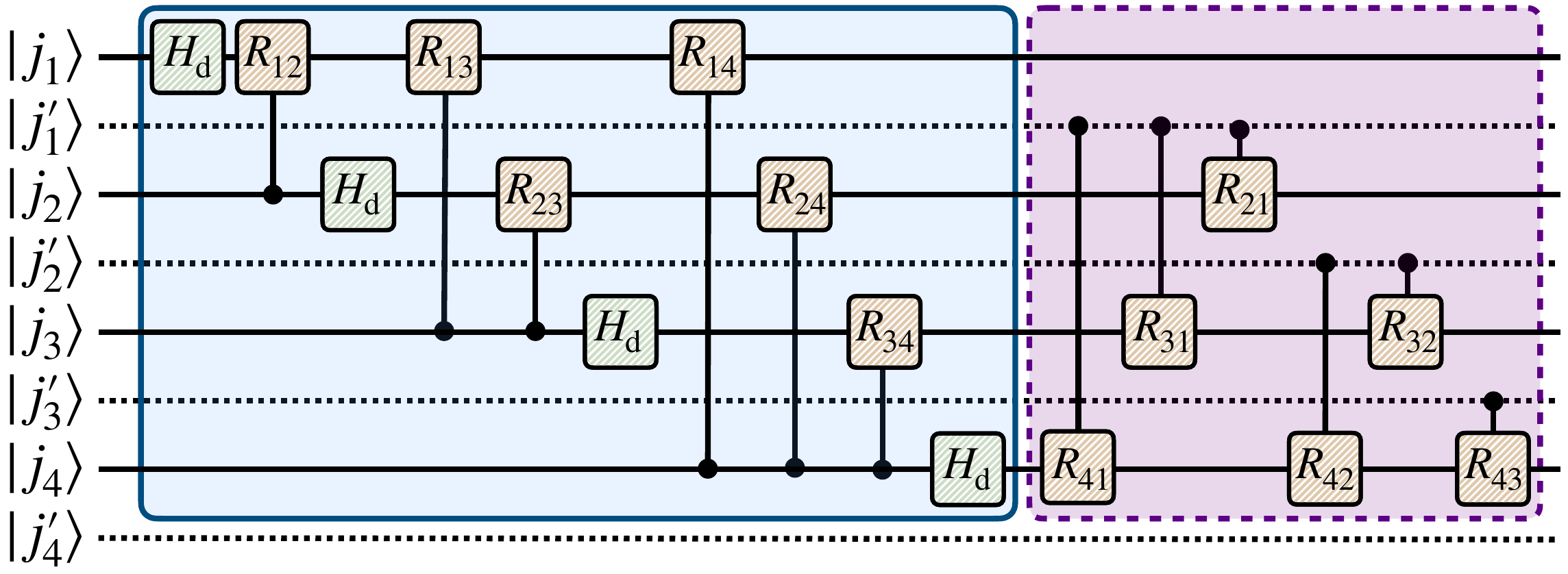}
    \caption{Circuit diagram of the damping transform $\DT$, which forms the non-unitary part of the $z$-transform $\ZT$. Each two-qubit controlled gate and the ``damping-Hadamard'' in this circuit will be mapped to a local tensor to obtain an MPO.}
    \label{fig:circuit}
\end{figure}

We follow the tensor-network reinterpretation of quantum circuits as matrix product operators, as used for the $\QFT$ in Ref.~\cite{chen2023quantum}, and apply it to the non-unitary damping circuit underlying our $z$-transform. The circuit for $\DT$ is shown in Fig.~\ref{fig:circuit}. Below we show how each gate in Fig.~\ref{fig:circuit} is mapped to a local tensor, so that the entire circuit can be written as a tensor network.

We start with the controlled damping gate
\begin{equation}
\raisebox{-2.0em}{\includegraphics[height=4.0em]{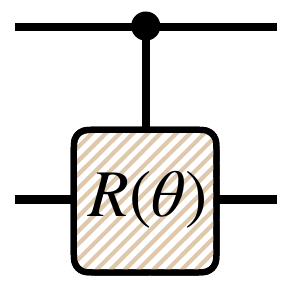}}
=
\ket{0}\!\bra{0} \otimes \openone
\;+\;
\ket{1}\!\bra{1} \otimes R(\theta),
\label{eq:cR}
\end{equation}
which applies the identity $\openone$ to the data register if the control is $\ket{0}$ and applies a non-unitary map \(R(\theta)=\mathrm{diag}(1,e^{-\theta})\) if the control is \(\ket{1}\).
This two-qubit operator can be factored into an inner product of two operator-valued vectors,
\begin{equation}
\ket{0}\!\bra{0} \otimes \openone
+
\ket{1}\!\bra{1} \otimes R(\theta)
=
\begin{pmatrix}
\ket{0}\!\bra{0} & \ket{1}\!\bra{1}
\end{pmatrix}
\begin{pmatrix}
\openone \\
R(\theta)
\end{pmatrix}.
\label{eq:factorization}
\end{equation}
This factorization rewrites the controlled damping gate as the contraction of two order-3 tensors, i.e., as a two-site MPO. The first tensor is the copy tensor,
\begin{equation}
\raisebox{-1.2em}{\includegraphics[height=2.5em]{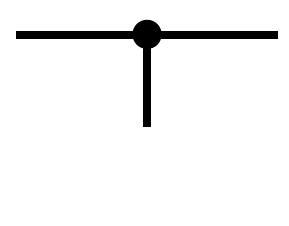}}
=
\begin{pmatrix}
\ket{0}\!\bra{0} & \ket{1}\!\bra{1}
\end{pmatrix},
\label{eq:copy_tensor}
\end{equation}
and the second is the damping tensor,
\begin{equation}
\raisebox{-1.2em}{\includegraphics[height=3.0em]{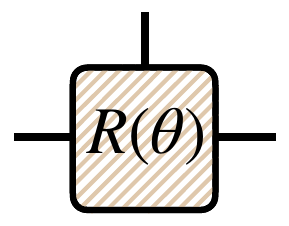}}
=
\begin{pmatrix}
\openone \\
R(\theta)
\end{pmatrix}.
\label{eq:damping_tensor}
\end{equation}
The copy tensor maps the control branches $\ket{0}$ and $\ket{1}$ onto a virtual bond index, and the damping tensor applies either $\openone$ or $R(\theta)$ conditioned on that index, so the control is carried by the MPO bond. Two such controlled-damping gates applied consecutively can be fused into a single MPO,
\begin{equation}
\raisebox{-3.5em}{\includegraphics[height=7em]{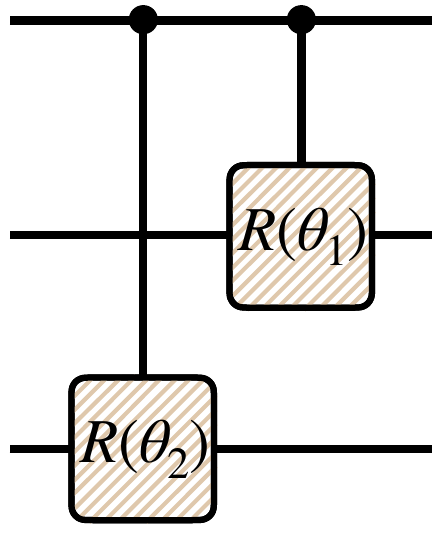}}
=
\raisebox{-3.5em}
{\includegraphics[height=7em]{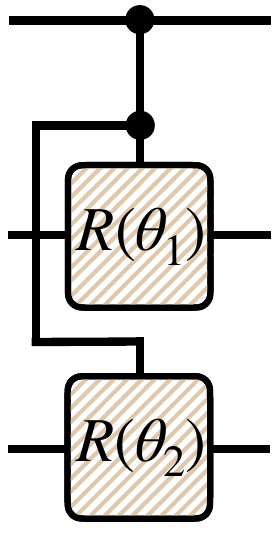}}
=
\raisebox{-3.5em}
{\includegraphics[height=7em]{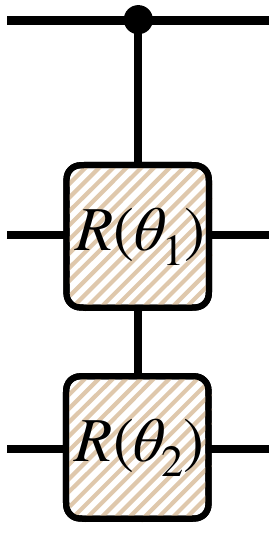}}
\label{eq:cRR_mpo}
\vspace{1mm}
\end{equation}
whose internal block is an order-4 tensor,
\begin{equation}
\raisebox{-1.7em}{\includegraphics[height=4.0em]{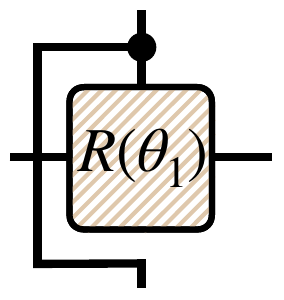}}
=
\raisebox{-1.7em}
{\includegraphics[height=4.0em]{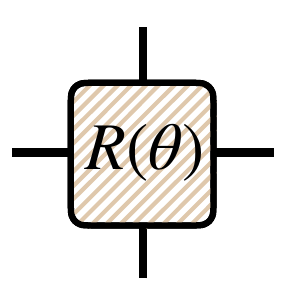}}
=
\begin{pmatrix}
\openone & 0 \\
0 & R(\theta)
\end{pmatrix}.
\label{eq:R4_block}
\end{equation}
Similarly, multiple damping gates sharing a common control can be fused into a controlled multi-site MPO; we refer to this as the `controlled-damping MPO'.

In addition, the on-site ``damping-Hadamard'' tensor is given by
\begin{equation}
\raisebox{-1.1em}{\includegraphics[height=2.2em]{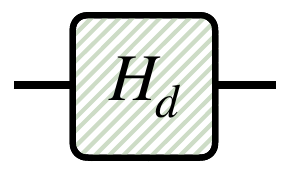}}
=
\frac{1}{\sqrt{2}}
\begin{pmatrix}
1 & 1 \\
1 & e^{-\omega_r/2}
\end{pmatrix}.
\label{eq:damped_hadamard}
\end{equation}
Using these tensor assignments, the damping circuit in Fig.~\ref{fig:circuit} maps directly to the MPO/tensor-network representation built from the damped-Hadamard and controlled-damping MPOs shown in Fig.~1 of the main text. We now outline the construction of the \emph{truncated} $\ZT$ MPO for this network.

\paragraph*{MPO contraction (zip-up / zip-down).}
We obtain the $\QFT$ MPO via the zip-up/zip-down scheme of Ref.~\cite{chen2023quantum}. For the damping network, we use a similar iterate-merge pattern with \emph{QR} preconditioning before truncation. 


A controlled-damping MPO is the multi-site MPO formed by fusing damping gates sharing a common control (Fig.~1 of the main text). For two adjacent controlled-damping MPOs acting on the single-register branch ($\ket{j}$): (i) locally contract the corresponding site tensors over physical indices; (ii) perform a bottom-to-top series of \emph{QR} factorizations to place the orthogonality center (OC) at site~1, writing $A=QR$, keeping $Q$ on-site and pushing $R$ upward--this canonizes the MPO; (iii) at the OC, take a truncated singular-value decomposition (SVD) $X=USV^\dagger$ with a relative cutoff, keeping $V$ on-site and pushing $US$ downward. Repeating the SVD/push cycle compresses and moves the OC to the bottom. The resulting MPO is then merged with the next controlled-damping MPO by the same steps. Iterating over all controlled-damping MPOs yields $\DT$ on $\ket{j}$.  We then repeat the same pairwise merge on the joint branch ($\ket{j}\!\otimes\!\ket{j'}$) to obtain the joint component; composing these two components gives the (single) $\DT$ MPO. Finally, composing the $\DT$ MPO with the $\QFT$ MPO, with appropriate QR sweeps and truncated SVDs, produces a single, compressed MPO realizing the full $z$-transform.

\begin{figure}[!t]
  \centering
  \includegraphics[width=0.98\columnwidth]{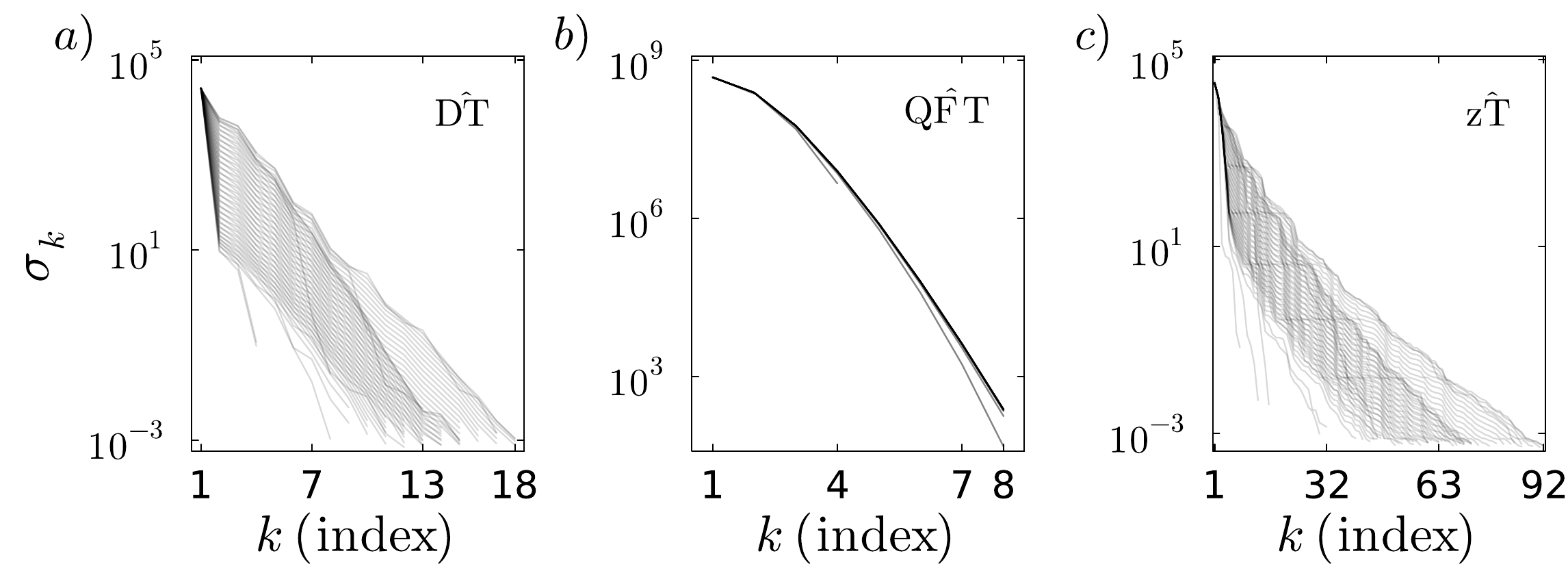}
  \caption{ Singular-value spectra across all internal bonds for $n=30$. Panels (a), (b), and (c) show $\sigma_k$ versus index $k$ for $\DT$, $\QFT$, and $\ZT$, respectively, on a $\log_{10}$ y-axis with relative cutoff $\tau=10^{-15}$ and $(\omega_r,\omega_i)=(2\pi,2\pi)$. Each curve corresponds to the spectrum at a single bond.}
  \label{fig:sv_all_bonds_n30}
\end{figure}

\begin{figure}[h!]
  \centering
  \includegraphics[width=0.98\columnwidth]{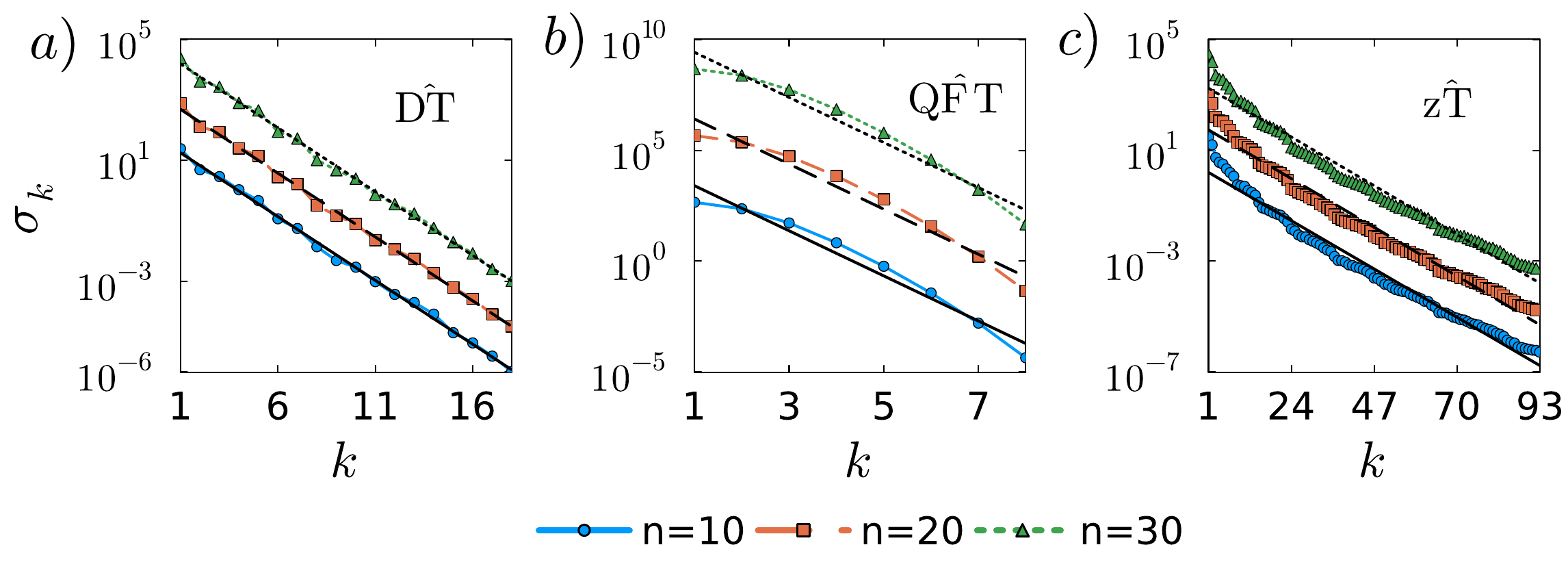}
  \caption{ Singular-value spectra at the first bond attaining the maximum bond dimension for $\DT$ (a), $\QFT$ (b), and $\ZT$ (c), for $n\in\{10,20,30\}$, are shown on a $\log_{10}$ y-axis with relative cutoff $\tau=10^{-15}$. Exponential fits of the form $\sigma_k = A e^{-\alpha k}$ (black lines), plotted on the same $\log_{10}$ scale, closely match the numerical data (markers), supporting an efficient low-rank MPO approximation.
  }
  \label{fig:sv_selected_bonds}
\end{figure}

The singular-value spectra $\sigma_k$ of the MPO decompositions are shown in Fig.~\ref{fig:sv_all_bonds_n30} for all internal bonds at $n=30$ ($\log_{10}$ y-axis). Figure~\ref{fig:sv_selected_bonds} focuses on the bond with the largest dimension and shows the spectra for $n\in\{10,20,30\}$; exponential fits, plotted on the same $\log_{10}$ scale, indicate an approximately exponential decay, $\sigma_k \propto e^{-\alpha k}$. 
This rapid decay supports an efficient low-rank MPO representation of the corresponding transforms, while the overall computational cost also depends on the compressibility of the input MPS.

\section{Input signals} \label{app:functions} 
\renewcommand{\thefigure}{D\arabic{figure}}
\setcounter{figure}{0}

For Fig.~2(c) of the main text, we use signals of length $N=2^{n}$ with samples indexed by $j=0,\dots,N-1$ and sampling interval $\Delta t = 5/2^{n}$. Random numbers are generated using Julia’s \texttt{MersenneTwister} pseudorandom generator (from the standard \texttt{Random} module) with fixed seeds to ensure exact reproducibility. The following signal families are used: (i) a pure sinusoid, $x[j] = \sin(2\pi j\,\Delta t)$; (ii) Gaussian noise, $x[j] \sim \mathcal{N}(0,1)$ independently for all $j$, generated with seed \texttt{1234}; (iii) a multi-sinusoid with exponential decay, $x[j] = \sum_{k=1}^{n_{\mathrm{terms}}} a_k \sin(\omega_k \Delta t\, j)\, e^{\lambda_k \Delta t\, j}$, with $n_{\mathrm{terms}} = 10$, amplitudes $a_k$ drawn uniformly from $[0,1]$ and $\ell_2$-normalized (seed \texttt{1001}), frequencies $\omega_k$ drawn uniformly from $[-20,20]$ (seed \texttt{2002}), and decay rates $\lambda_k$ drawn uniformly from $[-2,0]$ (seed \texttt{4004}), ensuring exponential decay; and (iv) a cusp-type signal, $x[j] = |\cos(2\pi j\,\Delta t)|^{0.8}$, which is non-smooth and used to test robustness against sharp features.

\section{Locating poles and zeros} \label{sec:pole_locations}
\renewcommand{\thefigure}{E\arabic{figure}}
\setcounter{figure}{0}

Although the finite-\(N\) transform has no true poles, the pole locations of the corresponding infinite series can still be estimated from the finite-$N$ data. Here, we consider the signal
\begin{equation}
    x_j\propto a^j\cos(\omega_0 j),
    \qquad j=0,\ldots,N-1,
\end{equation}
with $N=2^{20}$. The numerical parameters used in the calculation were
\begin{equation}
\begin{aligned}
    a&=1.00015002\,e^{\im 0.00204},\\
    \omega_0&=0.00612993.
\end{aligned}
\end{equation}
\bigskip
These parameters give the infinite-series pole locations
\begin{equation}
\begin{aligned}
    z_{+}&=0.99984164+0.00408931\im,\\
    z_{-}&=0.99981663-0.00816861\im.
\end{aligned}
\end{equation}
\bigskip

\begin{figure}
    \centering
    \includegraphics[width=1\linewidth]{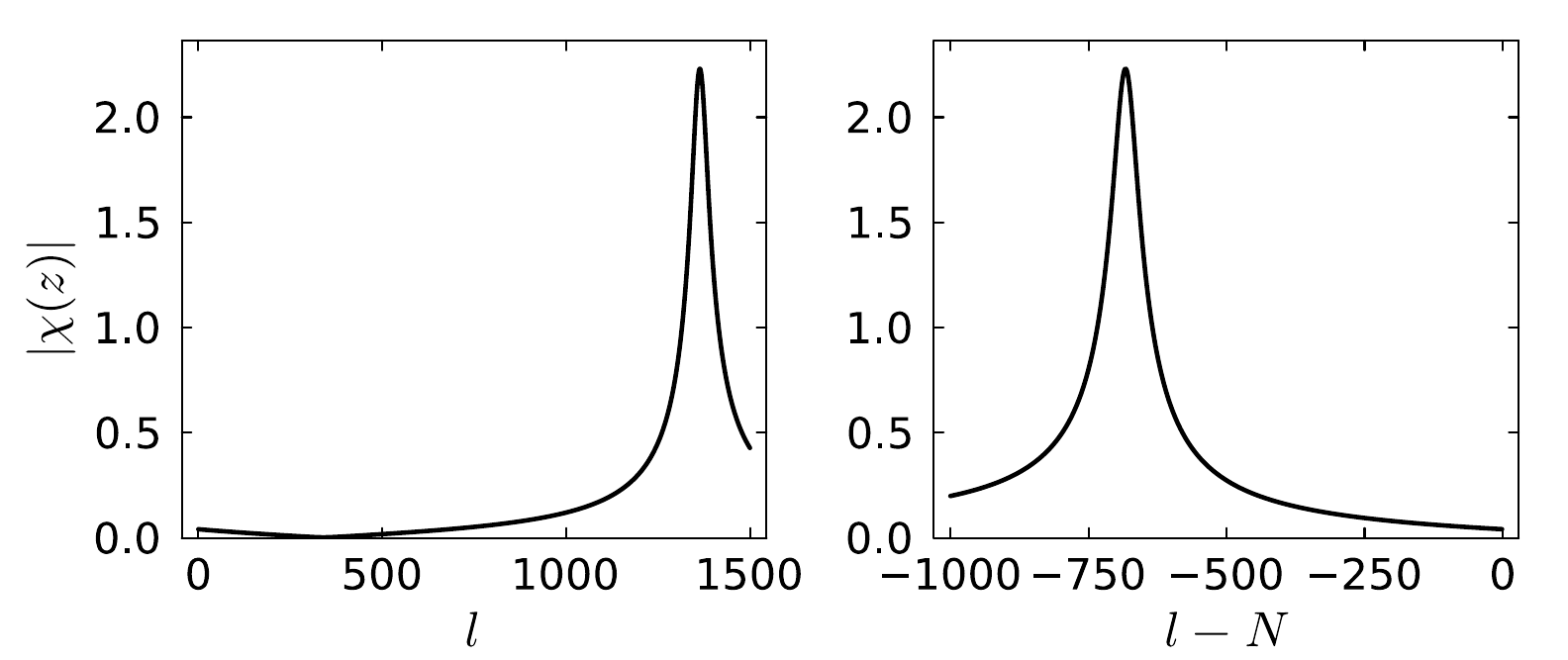}
    \caption{Unit-circle (\(k=0\)) angular slices in the two separated \(l\) regions. The dominant responses occur at \(l=1364\) and \(l=1047893\).}
    \label{fig:unit_circle_regions}
\end{figure}

\begin{figure}
    \centering
    \includegraphics[width=1\linewidth]{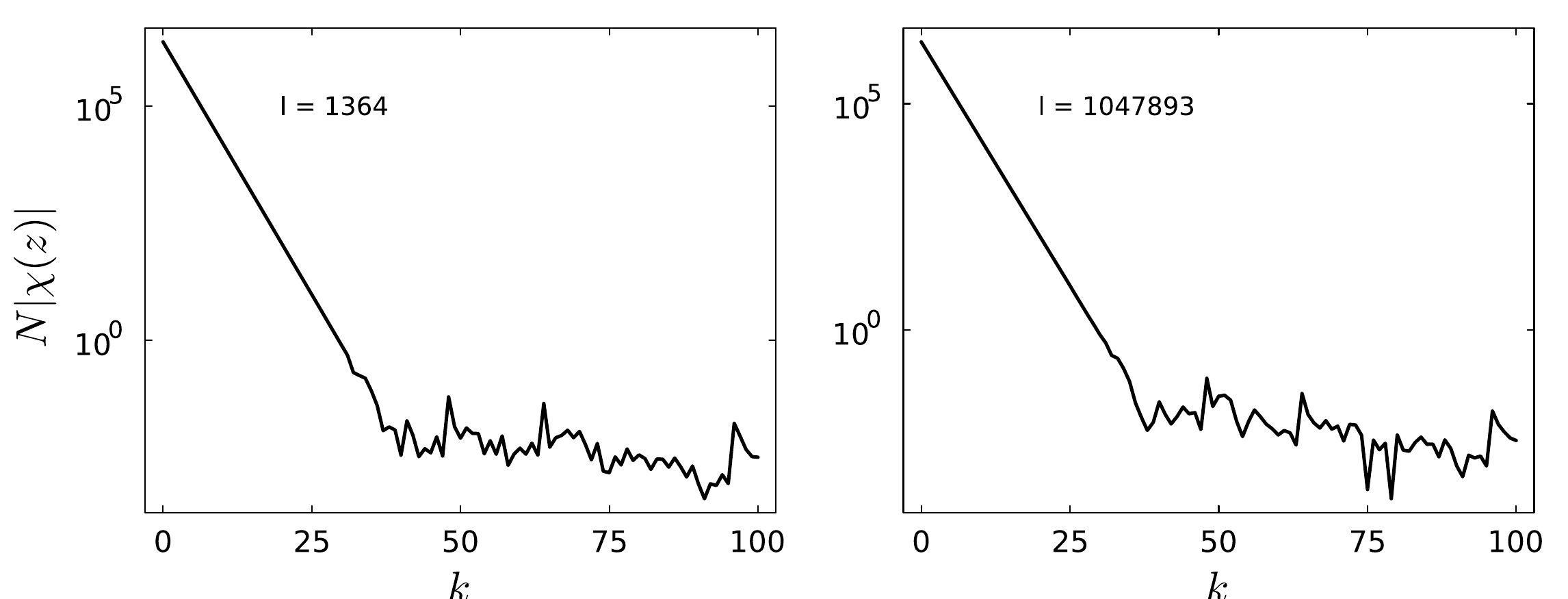}
    \caption{Radial slices of \(N|\chi(z)|\) at \(l=1364\) and \(l=1047893\). The pole is inferred from the onset of divergence-like growth, rather than from a resolved finite-\(N\) maximum.}
    \label{fig:radial_slices}
\end{figure}

We first examine the unit-circle slice (\(k=0\)), as shown in Fig.~\ref{fig:unit_circle_regions}. The dominant responses occur at  \(l=1364\) and \(l=1047893\), which identify the relevant spectral components. For each of these \(l\)-values, we then inspect the corresponding radial slice, shown in Fig.~\ref{fig:radial_slices}. Since the infinite-series transform diverges at the pole, the finite-\(N\) signature is not a resolved maximum but the onset of approximate exponential growth, visible as an approximately linear regime on the semilog plot. The corresponding \((k,l)\) values are then used to estimate the pole locations.

\subsection{Zero locations}
\label{sec:zero_locations}

\begin{figure}
    \centering
    \includegraphics[width=1\linewidth]{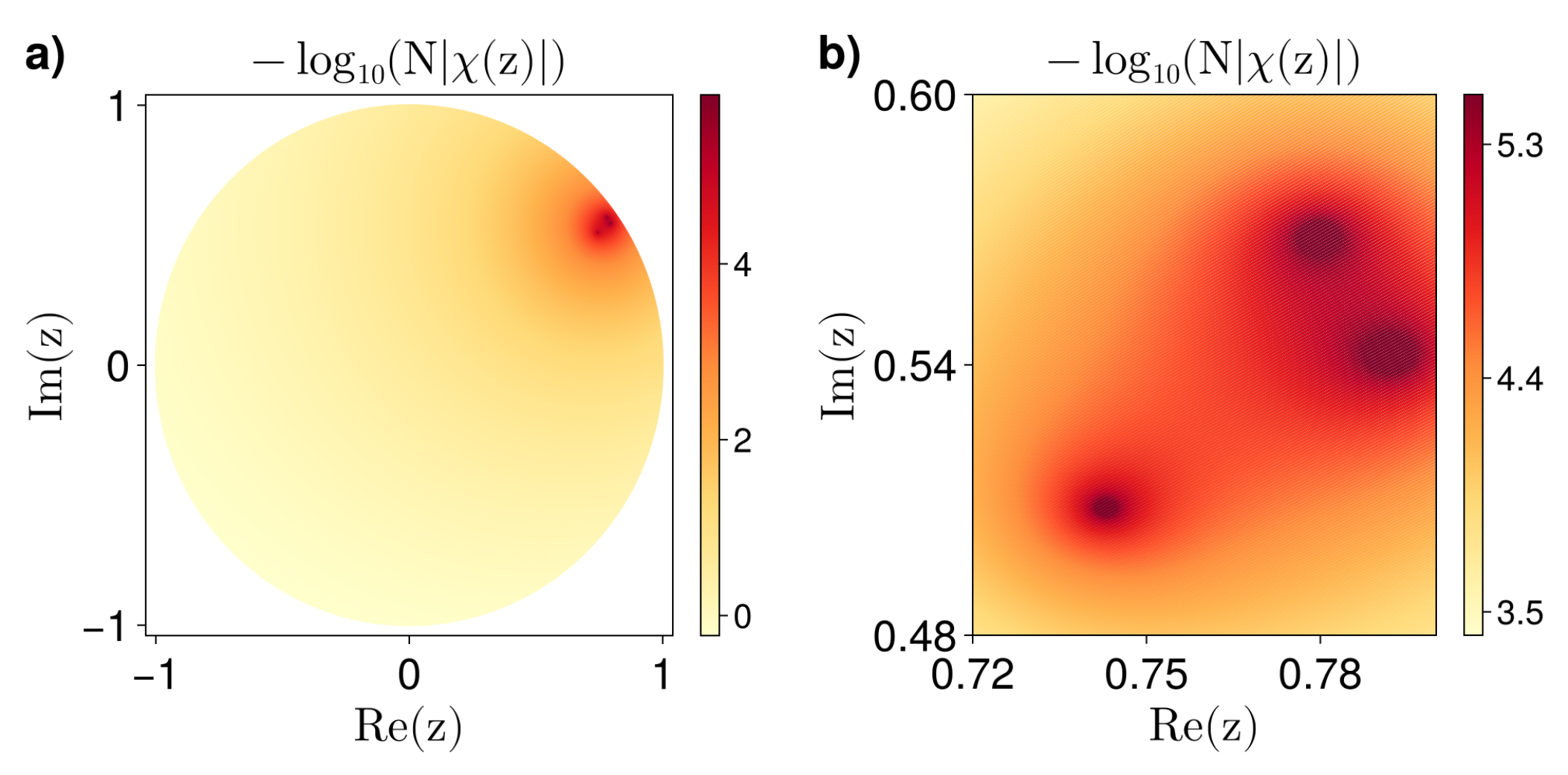}
    \caption{Zero finding for \(X(z)=e^{az}(z-\alpha)(z-\beta)(z-\gamma)\).
    (a) Coarse scan with \(\omega_r=2\pi\), showing the broader \(z\)-plane structure and the region containing the zeros.
    (b) Refined scan with \(\omega_r=0.5\) over the candidate region identified in (a), where the three nearby zero locations are inferred as distinct localized maxima in \(-\log_{10}\!\bigl(N|\chi(z)|\bigr)\).}
    \label{fig:zero_scan_three_roots}
\end{figure}

The same approach can also be used to identify zeros of the $z$-transform. As a test case, we consider
\begin{equation}
X(z)=e^{az}(z-\alpha)(z-\beta)(z-\gamma),
\end{equation}
with \(\alpha = 0.90\,e^{\im 0.60},\;
\beta  = 0.96\,e^{\im 0.60},\;
\gamma = 0.965\,e^{\im 0.63}\), and \(a=0.4\).
These zeros are chosen to lie close to one another, in order to test the resolution of our method. For the full analytic function they are exact zeros, while in the finite-$N$ truncated transform they appear numerically as localized minima of $|X(z)|$. As shown in Fig.~\ref{fig:zero_scan_three_roots}, a coarse scan first identifies the relevant region of the \(z\)-plane, after which a refined local scan resolves the three nearby zero locations and estimates their positions as \(\alpha = 0.90001\,e^{\im 0.59994},\;
\beta  = 0.96010\,e^{\im 0.59993},\;
\gamma = 0.96490\,e^{\im 0.63006}\).

%